\newif\ifsingle

\ifsingle
\documentclass[11pt,draftclsnofoot, onecolumn]{IEEEtran}		
\else		
\documentclass[10pt,final, twocolumn]{IEEEtran}
\fi

\usepackage{microtype}
\usepackage{graphicx} 
\usepackage{booktabs} 

\usepackage{amsmath,graphicx}
\usepackage{amssymb}
\usepackage{acronym,cite}
\usepackage{caption}
\usepackage{makecell}
\usepackage{multirow}
\usepackage{standalone}
\usepackage{subcaption}
\usepackage{tikz}
\usepackage{verbatim}
\usepackage[linesnumbered,lined,boxed,commentsnumbered]{algorithm2e}
\usepackage[normalem]{ulem}

\usepackage{damusic}

\usetikzlibrary{trees,arrows}

\usepackage[bookmarks,colorlinks]{hyperref}

\newcommand{\add}[1]{#1}

\newcommand{\remove}[1]{\iffalse{#1}\fi}

\newcommand{\subsubsubsection}[1]{{\bf #1:}}

\IEEEoverridecommandlockouts

\title{DA-MUSIC: Data-Driven DoA Estimation via Deep Augmented MUSIC Algorithm}

\author{
\IEEEauthorblockN{Julian P. Merkofer,~\IEEEmembership{Student Member, IEEE}, Guy Revach,~\IEEEmembership{Student Member, IEEE}, Nir Shlezinger,~\IEEEmembership{Member, IEEE}, Tirza Routtenberg,~\IEEEmembership{Senior Member, IEEE}, and Ruud J. G. van Sloun,~\IEEEmembership{Member, IEEE}\\}
\thanks{Parts of this work were presented in the 2022 IEEE International Conference on Acoustics, Speech, and Signal Processing (ICASSP) as the paper~\cite{merkofer2021deepAugMUSIC}.
J. P. Merkofer is with the EE Dpt., Eindhoven University of Technology, Eindhoven,  The Netherlands (e-mail: j.p.merkofer@tue.nl), this work was partially done while being with the ETH Zürich.
G. Revach is with the D-ITET, ETH Zürich, Switzerland, (email: grevach@ethz.ch).
N. Shlezinger and T. Routtenberg are with the School of ECE, Ben-Gurion University of the Negev, Beer Sheva, Israel (e-mail: \{nirshl; tirzar\}@bgu.ac.il). T. Routtenberg is also with the ECE Dpt., Princeton University, Princeton, NJ, United States. 
R. J. G. van Sloun is with the EE Dpt., Eindhoven University of Technology, and with Phillips Research, Eindhoven,  The Netherlands (e-mail: r.j.g.v.sloun@tue.nl).
}

\vspace{-0.7cm}
}

\begin{document}

\maketitle
\pagestyle{plain}
\thispagestyle{plain}

\begin{abstract} 
\Ac{doa} estimation of multiple signals is pivotal in sensor array signal processing. 
A popular multi-signal \ac{doa} estimation method is the \ac{music} algorithm, which enables high-performance super-resolution \ac{doa} recovery while being highly applicable in practice. \ac{music} is a \acl{mb} algorithm, relying on an accurate mathematical description of the relationship between the signals and the measurements and assumptions on the signals themselves (non-coherent, narrowband sources). As such, it is sensitive to model imperfections. 
In this work we propose to overcome these limitations of \ac{music} by augmenting the algorithm with specifically designed neural architectures. Our proposed \ac{damusic} algorithm is thus a hybrid \acl{mb}/\acl{dd} \ac{doa} estimator, which leverages data to improve performance and robustness while preserving the interpretable flow of the classic method.  
\ac{damusic} is shown to learn to overcome limitations of the purely \acl{mb} method, such as its inability to successfully localize coherent sources as well as estimate the number of coherent signal sources present. We further demonstrate the superior resolution of the \ac{damusic} algorithm in synthetic narrowband and broadband scenarios as well as with real-world data of \ac{doa} estimation from seismic signals.
\end{abstract}

\acresetall

\section{Introduction} \label{sec:intro}

Source separation, localization, and tracking are crucial tasks in sensor array  processing. In particular, \acf{doa} estimation of multiple, broadband, and possibly coherent signal sources plays a key role in a wide range of applications, including radar, communications, image analysis, and speech enhancement \cite{bilik2019riseOfRadar, rahamim2004sourceLoc, foutz2008narrowband}. Over the last decades, a multitude of different \ac{doa} estimation algorithms have been proposed, and the problem is an active area of research \cite{krim1996twodecades, ahmad2014threedoa}. 

A leading scheme employed in many \ac{doa} applications is the popular \acf{music} algorithm~\cite{schmidt1986music}, which can provide asymptotically unbiased estimates of the number of incident wavefronts present, their approximate frequencies, and their \acp{doa}. \Ac{music} and other classic approaches, e.g., conventional beamforming~\cite{bartlett1948beamformer} and \acs{mvdr} beamforming~\cite{capon1969mvdrbf}, are based on knowledge of the underlying statistical model; this dependency induces several key limitations. Among the drawbacks of the \acf{mb} approaches is the inherent limitation of the signal model from which they are derived, which considers narrowband signals. This results, for example, in the inability to consistently estimate the \ac{doa} of correlated (coherent) signals, as well as a failure to resolve closely spaced signals with an insufficient number of samples or a low \ac{snr} \cite{schmidt1986music, krim1996twodecades}.

To extend narrowband models to broadband \ac{doa} estimation, various extensions and alternative approaches have been explored~\cite{tun2009cha4}, because when the impinging signals are broadband, multiple frequency ranges can carry different information regarding the \ac{doa} angles at hand. Generally, these extensions from narrowband to broadband can be subdivided into coherent and incoherent processing methods.
In coherent processing methods, the covariance matrices of the observations in different frequency bins are coherently combined using certain transformation matrices. Most coherent techniques are based on the \ac{css} concept~\cite{wang1985coherent}, which focuses the transformations of the covariances at different frequencies into a surrogate narrowband model, yet utilizes different transformation matrices. Incoherent broadband processing  methods  combine \ac{doa} estimates obtained separately for each frequency bin \cite{su1983thesignal}.

The recent success of \acf{dd} \ac{dl} across a wide range of disciplines gave rise to \ac{nn}-aided \ac{doa} estimators~\cite{grumiaux2022survey}. The works \cite{chakrabarty2019multi,liu2018direction,Wu2019DeepCN,hammer2021dynamically} implemented model-agnostic \ac{doa} estimation using a dense \ac{nn}, a \ac{cnn}, \add{a sparse-connected \ac{cnn},} and a U-Net architecture, respectively. \add{Other methods \cite{Papageorgiou2020DeepNF, Barthelme2020AML} utilize the knowledge of the spatial covariance matrices, where \cite{Papageorgiou2020DeepNF} trained a \ac{cnn} based on a classification task (allowing it to also operate with unknown number of sources) and \cite{Barthelme2020AML} analyzed purely \ac{dd} estimators and \ac{dd} methods in combination with \ac{mle}. The works \cite{Hu2020LowComplexityDD, Oliveira2022DeepMLEFB} also combined \ac{dl} with \ac{mle} by using multiple dense \acp{nn} and a ResNet, respectively, to estimate a subset of candidate angles. \cite{LimadeOliveira2022ResNetAF} extends the architecture of \cite{Oliveira2022DeepMLEFB} to estimate the number of sources with the ResNet.} While such black-box \acp{nn} enable handling array imperfections due to their model-agnostic nature, they involve highly parameterized models that may be computationally intensive and lack the interpretability  of \ac{mb} methods. 

Alternatively, \acp{nn} were used to robustify the \ac{mb} \ac{music} as a form of hybrid \ac{mb}/\ac{dd} system \cite{shlezinger2020model,shlezinger2022model}. Specifically, the work \cite{elbir2020deepmusic}\add{and \cite{Liu2021SingleSD}} proposed to estimate the discretized \ac{music} spectrum from the \add{(spatially smoothed)} covariance matrix of the measurements through the utilization of multiple convolutional \acp{nn}. While \add{these}\remove{this} method\add{s are}\remove{is} more robust to model inaccuracies compared with the original \ac{music} algorithm, utilizing the \ac{music} spectrum as a label for training causes \add{them}\remove{it} to experience the same drawbacks as \add{their}\remove{its} \ac{mb} counterpart. 
Another \ac{dd} approach proposed in \cite{barthelme2021doa} considered systems with subarray sampling and uses \acp{nn} to obtain a single estimated covariance matrix from incoherent subarray measurements. This \ac{nn}-aided estimate is utilized for \ac{doa} recovery via the subspace-based \ac{music} algorithm. The method addresses the fundamental dependency of \ac{music} on the estimated covariance matrix, thereby robustifying the \ac{music} algorithm, yet it does not fully exploit the \acp{nn}' ability to improve \ac{music}, as the \ac{nn} is trained using the true covariance matrix as a label, without considering its downstream task. 
\add{Furthermore, the work \cite{Hoang2022DeepLC} proposed a hybrid \ac{mb}/\ac{dd} approach that includes an \ac{evd} of \ac{ftmr} matrices~\cite{Zhang2019MultipleToeplitzMR} and classification deep \acp{nn} for estimation of a \ac{music}-like spectrum and detection of number of sources present via eigenvectors and eigenvalues}

The works \cite{chakrabarty2017bbdoacnn,zhu2019deeplearnbb} applied \acp{cnn}  for broadband \ac{doa} estimation of a single source. The two approaches, however, utilize very different preprocessing methods and are trained as classification and regression tasks, respectively. 
While the classification approach limits itself to a fixed resolution, the regressor of \cite{zhu2019deeplearnbb} outputs the \ac{doa} angles directly. Therefore, it can achieve an arbitrary resolution, yet it is limited to the spatial covariance matrix as input. \add{Another approach \cite{Wu2019CoherentSL} first decomposed the broadband signals into non-overlapping narrowband components and used \acl{svr} to obtain the \ac{doa}.}

The\remove{se} limitations of both existing \ac{mb} and \ac{dd} \ac{doa} estimation algorithms in resolving multiple signals that are possibly broadband and coherent motivate the derivation of a \ac{nn}-aided \ac{doa} estimator capable of leveraging data to enable \ac{music} to successfully operate in such scenarios.

\remove{In this work we propose}\add{We introduced} \ac{damusic} \add{in} \cite{merkofer2021deepAugMUSIC}, a hybrid \ac{mb}/\ac{dd} implementation of \ac{music}, which exploits the structure of the classic \remove{\ac{music}}\add{subspace} algorithm while augmenting it with \acp{nn} to learn to enhance its performance. 
The proposed architecture overcomes the fundamental limitations of \ac{music}, enabling it to\remove{ successfully estimate the number of signal sources present as well as} accurately detect the\remove{ir} locations \remove{for}\add{of} coherent sources\remove{ in narrowband as well as broadband scenarios}. Our design builds upon the insight that the sensitivity to model mismatch and inability to handle coherent and broadband sources of \ac{music} lie in its estimation of the noise and signal subspaces through an \ac{evd} of the empirical covariance matrix. Accordingly, the proposed \ac{damusic} improves this crucial step by obtaining a surrogate, pseudo covariance matrix through a \ac{rnn} from the measurements, which is learned along with a \ac{nn} that acts as a peak finder.
\add{This work presents crucial extensions to \ac{damusic} allowing it to operate with an unknown and varying number of signal sources by successfully estimating the number of sources for non-coherent and coherent signals. An additional neural augmentation allows the categorization into noise and signal eigenvectors to be obtained in a learned fashion by training a classifier. Numerical results show an estimation accuracy of 98\% in comparison to \ac{music} with 89\% accuracy in the same non-coherent scenario. We further demonstrate \ac{damusic}'s capabilities in various \ac{snr} domains, three different broadband scenarios, as well as real-world seismic signals. \Ac{damusic} not only manages to focus frequency components but is also able to concurrently focus an interdependent elevation angle to receive a stable azimuth estimation.} 


\remove{Our experimental study demonstrates the ability of the proposed architecture to notably improve upon both \ac{mb} and \ac{dd} \ac{doa} estimators. In particular, we show that the \ac{damusic} algorithm outperforms its unaltered version in localization accuracy and resolution for both synthetic settings, as well as for \ac{doa} estimation from measurements of complex broadband seismic signals.}

The rest of the paper is organized as follows: Section \ref{sec:sysModel} describes the assumed system model and surveys some of the related work of \ac{doa} estimation, as well as the technical details of the \ac{mb} \ac{music} algorithm; Section \ref{sec:da-music} introduces the \ac{damusic} algorithm; Section \ref{sec:numEvals} presents the results of the simulations; and Section \ref{sec:conclusion} provides concluding remarks.

Throughout the paper, we use boldface lower-case letters for vectors; e.g., $\vect{x}$, and boldface uppercase letters, e.g., for $\mat{X}$. The $i$th entry of a vector $\vect{x}$ is denoted by $x_i$ \add{and entries separated  by commas represent column vectors.} We use calligraphic letters to denote the discrete-time Fourier transform, e.g., $\ft{X}(\omega)$ is the frequency representation of a signal $x(t)$. We use 
$(\cdot)^\transpose$ and $(\cdot)^\hermit$ to denote 
the transpose and Hermitian operators respectively, and $\norm{\cdot}$ and $\norm{\cdot}_\forb$ to denote the $\ell_2$ and Forbenius norms respectively. Further, $\mathcal{U}$ and $\mathcal{CN}$ represent the uniform and the complex normal distributions. The symbol~$\mathbb{I}$ denotes the identity matrix.
%

\section{System Model and Preliminaries} \label{sec:sysModel}
In this section, we detail the system model for which we develop the proposed \ac{damusic} algorithm in Section~\ref{sec:da-music}. To that aim, we first discuss the signal model in Subsection~\ref{ssec:Signal}.
Then,  we formulate the \ac{doa} estimation problem in Subsection~\ref{sec:problem}, and survey some of the related literature in Subsection~\ref{sec:literature}. Since our proposed solution builds upon the \ac{music} algorithm, 
we recall \ac{music} in Subsection~\ref{sec:music} and briefly discuss the \ac{css} method for broadband extension in Subsection~\ref{sec:coherentBB}.

\subsection{Signal Model}\label{ssec:Signal}
We distinguish the considered signal model into two cases: the conventional narrowband setting and its more general broadband setup. For more details regarding the two models, we refer the reader to the textbooks \cite{friedlander2009cha1, tun2009cha4}, and \cite{yu2010introdoa}.

\subsubsection{Narrowband} \label{sec:nbSys}
The most common setup in the  array signal processing literature concerning \ac{doa} estimation is the narrowband setting. Here,  the time it takes for the waves to propagate the array is assumed to be negligible such that the occurring delays are sufficiently small. Therefore, the signal measurement model for an arbitrary array structure consisting of $M$ sensor elements measuring $D$ impinging narrowband signals takes the following form:
\begin{equation} \label{eq:sysmod}
    \vect{x}(t) = \mat{A}(\boldsymbol{\theta}) \ \vect{s}(t) + \vect{v}(t).
\end{equation}
In \eqref{eq:sysmod}, the measurements $\vect{x}(t) \in \cmplx^M$ at time instance $t$ depend on the signals $\vect{s}(t)  \in \cmplx^D$, which originate from the unknown angles $\boldsymbol{\theta}=[\theta_1,\ldots,\theta_D]$, while $\vect{v}(t) \in \cmplx^M$ is additive white Gaussian noise. The matrix $\mat{A}(\boldsymbol{\theta}) \in \cmplx^{M \times D}$ contains the steering vectors $\{\vect{a}(\theta_d)\}$, i.e.,
\begin{equation}
    \mat{A}(\boldsymbol{\theta}) = \begin{bmatrix}\vect{a}(\theta_1) \dots \vect{a}(\theta_D)\end{bmatrix},
\end{equation}
where $\{\theta_d\}$ are the source directions.
For example, the steering vector $\vect{a}(\psi)$ for a \ac{ula} with an element spacing of $\Delta m = \ell / 2$, where $\ell$ is the wavelength of the signals, is defined for direction $\psi$ as 
\begin{equation}
    \vect{a}(\psi) = \begin{bmatrix}1, \ e^{-j \pi \sin{\psi}}, \ \dots, \ e^{-j \pi (M-1) \sin{\psi}}\end{bmatrix}.
\end{equation}
Consequently, the steering vectors (and, in turn, the matrix $\mat{A}(\boldsymbol{\theta})$) specify the underlying array geometry.
The collection of the measurements at the array elements over multiple time instances is defined as 
\begin{equation}
\label{eqn:XMat}
     \mat{X}= \begin{bmatrix}\vect{x}(1) \ \dots \ \vect{x}(T)\end{bmatrix},
\end{equation}
where $T$ is referred to as the number of snapshots.

\subsubsection{Broadband} \label{sec:bbSys}
In practice, many signals are broadband, and the delay caused by propagating the array aperture needs to be incorporated into the signal model. The following notation models the measurements received at array element $m \in \{1, ..., M\}$
\begin{equation} \label{eq:bbsysmod}
    x_m(t) = \sum_{d=1}^D s_d(t - \tau_{md}) + v_m(t),
\end{equation}
where $\tau_{md}$ represents the time delay of the $m$th array element measuring source~$d \in \{1, ...., D\}$ compared to the measurement at the reference element $m=1$.
Transformed to the frequency domain, the relationship in~\eqref{eq:bbsysmod} at frequency $\omega$ becomes
\begin{equation} \label{eq:bbsysmodfreq}
    \ft{X}_m(\omega) = \sum_{d=1}^D e^{-\imag \omega \tau_{md}} \ \ft{S}_d(\omega) + \ft{V}_m(\omega),
\end{equation}
which can in turn be written in vector-matrix form analogous to the narrowband system model~\eqref{eq:sysmod} as follows:
\begin{equation} \label{eq:bbsysmodfreqVM}
    \ft{X}(\omega) = \mat{A}(\omega, \boldsymbol{\theta}) \ \ft{S}(\omega) + \ft{V}(\omega).
\end{equation} 
For a \ac{ula}, 
the broadband steering vectors are defined for frequency $\omega$ and angle of interest $\psi$ as
\begin{equation}
    \vect{a}(\omega,\psi) = \begin{bmatrix}1, \ e^{- \imag \omega \frac{\Delta m}{c} \sin{\psi}}, \ \dots, \ e^{- \imag \omega (M-1) \frac{\Delta m}{c}\sin{\psi}}\end{bmatrix},
\end{equation}
where $c$ is the propagation velocity.

\begin{figure}
    \centering
    \includegraphics[width=0.9\columnwidth]{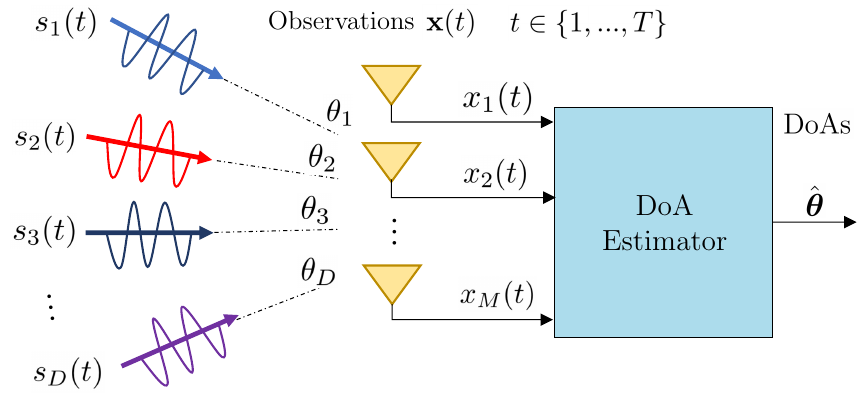}
    \caption{\ac{doa} estimation illustration.}
    \label{fig:DOAest}
\end{figure}

\subsection{Problem Formulation} \label{sec:problem}
\Ac{doa} estimation is concerned with localizing signal sources by determining their angles of incidence utilizing the measurements of an array aperture \cite{friedlander2009cha1}. Formally, this corresponds to estimating the angles $\boldsymbol{\theta} = [\theta_1, \dots, \theta_D]$ from the measurements $\vect{x}(t) = [x_1(t), \dots, x_M(t)]$  
measured at multiple time instances $t \in \{1, ..., T\}$. An illustration of such a system is depicted in Fig.~\ref{fig:DOAest}.
The following additional assumptions are imposed upon the observation model as well as the derived synthetic data. The signals are uncorrelated to the noise (though signals may possibly be correlated with each other)  and generated in the far-field region. And there is uniform propagation in all directions in an isotropic and non-dispersive medium.
We further assume that we have knowledge (though possibly mismatched) of the underlying array geometry, implying that we can compute an approximation of $\vect{a}(\psi)$ or $\vect{a}(\omega, \psi)$.
We consider a data-driven setting where we have access to  training data. 
 This data is a set of $U$ pairs, $\{(\mat{X}_u, 
 \boldsymbol{\theta}_u)\}_{u=1}^{U}$, each comprising the observations 
 and \ac{doa} angles from where the signals originated. In many scenarios; e.g., in  wireless communications, a specific training set can be developed before deployment. If the ground-truth \acp{doa} are not obtainable at all, the data-driven \ac{doa} estimator must be trained by utilizing synthetic data that closely describes the real signals.
Our goal is thus to leverage the available domain knowledge and data to design a system for recovering the \acp{doa} $\boldsymbol{\theta}$ from a corresponding observation matrix $\mat{X}$ whose columns are $T$ snapshots of the measured waveforms at the $M$ sensors. 

\begin{figure*}
\centering
\includegraphics[width=0.8\linewidth]{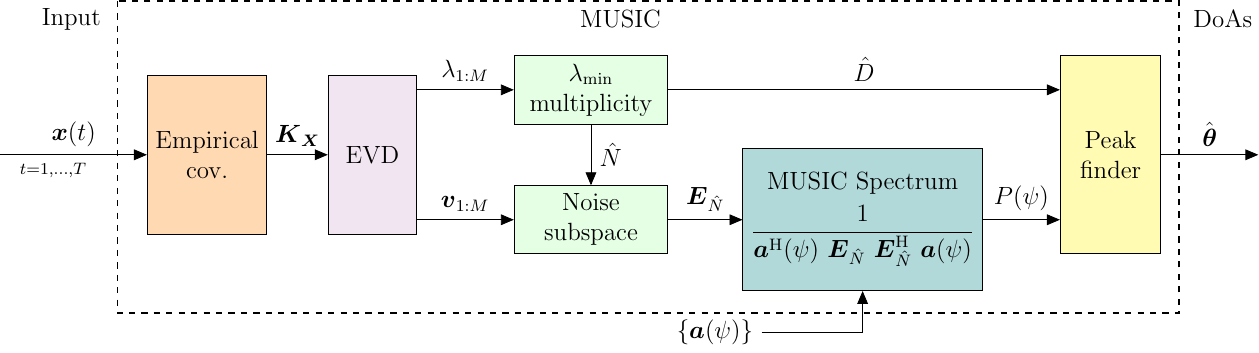}
\caption{Block diagram of the \ac{music} algorithm.}
\label{fig:bdMUSIC}
\end{figure*}

\subsection{Related Literature} \label{sec:literature}
We next provide an overview of relevant \ac{doa} estimation methods based on the above problem formulation, motivating the need for our proposed \ac{damusic} detailed in Section~\ref{sec:da-music}. An extensive overview of various \ac{mb} \ac{doa} estimators can be found in \cite{krim1996twodecades, ahmad2014threedoa}, and a recent literature review of \ac{dd} \ac{doa} estimation approaches can be found in \cite{ge2021dlindoa}. We thus only discuss some representative \ac{mb} methods, followed by reviewing relevant broadband extensions, and conclude with \ac{dd} architectures for narrowband as well as broadband signals.

\subsubsection{\ac{mb} Narrowband Estimators}
The conventional beamformer (i.e. maximizing the steered response) is a basic approach to \ac{doa} estimation and an extension of classical Fourier-based spectral analysis \cite{bartlett1948beamformer}. Various improvements and alternative beamforming methods have been developed, such as the \acl{mvdr} beamformer \cite{capon1969mvdrbf} and other adaptive beamformers \cite{frost1972adaptive}. An overview of  beamforming techniques can be found in \cite{veen1988beamforming}.

An alternative family of \ac{doa} estimators is based on subspace methods, which aim at recovering the \acp{doa} by identifying the noise and signal subspaces. The \ac{music} algorithm \cite{schmidt1986music} is a highly popular subspace-based method and has been researched extensively. Being the focus of this paper, more information and a detailed description of the algorithm can be found in Section~\ref{sec:music}. Extensions of \ac{music} include Root-MUSIC \cite{friedlander1993rootMUSIC}, a polynomial-rooting version, as well as spatially smoothed \ac{music}, which removes the correlation between the incident signals by dividing the receiver array into overlapping subarrays \cite{shan1985spatialsmoothing}. In practice, however, the number of coherent sources is mostly unknown, and therefore the decorrelation effect of spatial smoothing is not obvious \cite{chen2011onspatialsmoothing}. Another popular subspace-based method for \ac{doa} estimation is  \ac{esprit}  \cite{paulraj1985esprit} and its variations \cite{volodymyr2009doaesprit}. These methods rely heavily on the accuracy of the underlying model assumptions, are generally sensitive to array aperture perturbations, and are inherently derived for narrowband signals.

\subsubsection{\ac{mb} Broadband Estimators}
Broadband \ac{doa} estimation algorithms can be categorized into two groups: incoherent methods and coherent methods. Generally, incoherent methods use \acp{ifb} to process the \ac{doa} information at every frequency separately and then combine the results of these narrowband \ac{doa} estimations. There are many different  variations in the implementation of this approach \cite{wax1984spatio, chan1999doaest, arg2007bbvarofmusic}, which typically vary in the computation of the covariance matrices. Unfortunately, the computational complexity increases with each frequency bin.

Conversely, coherent methods combine the covariance matrices estimated for \acp{ifb} in order to apply narrowband techniques directly over a single, so-called focused covariance matrix. An overview of coherent techniques can be found in \cite{tun2009cha4,yoon2006doaest}. A leading approach for coherent broadband \ac{doa} estimation is based on the \ac{css} method \cite{wang1985coherent}, with example estimators given in \cite{krolik1990focused, ta-sung1994effwideband, fried1991dfforwide}. Many variations of the \ac{css} method are concerned with the crucial aspect of the focusing strategy, proposing different techniques to combine the covariances at each \ac{ifb} to obtain a faithful narrowband formulation. See, for example, \cite{claudio2001waves,yoon2006tops} and the more recent summary~\cite{ma2019wideband}. These methods, however, experience significant bias with larger angular sectors of interest and can impose additional model assumptions such as noise statistics.

\subsubsection{\ac{dd} Estimators}
Inspired by the dramatic success of \ac{dl} in computer vision and natural language processing, recent years have witnessed a growing interest in the application of \acp{nn} for \ac{dd} \ac{doa} estimation. \add{A recent review of \ac{dl} \ac{doa} estimation approaches can be found in \cite{Ge2021DeepLA} and essential methods were covered in Section \ref{sec:intro}. \Ac{dl} architectures presented in the numerical evaluations or otherwise closely related to this work are summarized and discussed below.}

\add{The work \cite{Papageorgiou2020DeepNF} implemented a \ac{cnn} architecture for \ac{dd} \ac{doa} estimation based on a multi-label classification task. Particularly, the method takes real, imaginary, and phase information of the sample covariance as a three-channel input and predicts a probability grid of directions. Given the binary cross-entropy loss for training, the method is capable of operating with a potentially unknown and varying number of sources. However, with a growing number of sources, the number of classes grows exponentially.}

\add{Similarly to the above, the methods in \cite{elbir2020deepmusic, Liu2021SingleSD} propose \acp{cnn} for \ac{doa} estimation by taking the same channels of the sample covariance matrix as input (spatially smoothed covariance of single snapshot for \cite{Liu2021SingleSD}). However, the \acp{cnn} are trained as a regression task estimating a discretized segment of the \ac{music} spectrum. Thereby, they not only inherit the drawbacks and imitations of \ac{music} but are, besides increased robustness, dependant on the underlying model assumptions.}

\add{The sample covariance matrix estimate deviates from the actual covariance, especially with multiple possibly coherent broadband signals which affect \ac{mb} as well as \ac{dd} performance. The work \cite{barthelme2021doa} considered systems with subarray sampling and trained a \ac{nn} for covariance matrix reconstruction. The method takes subarray covariance matrices as input and predicts a full covariance matrix. This \ac{nn}-aided estimate is then utilized for \ac{doa} recovery with the \ac{music} algorithm. While this approach addresses the fundamental dependency of \ac{music} and other \add{doa} estimators on the estimated covariance matrix, it is still dependent on their subarray estimate, and cannot further influence or react to \ac{music}'s performance.}

\add{The approach presented in \cite{Hoang2022DeepLC} proposed a hybrid \ac{mb}/\ac{dd} framework with dense \acp{nn} for localization and estimation of the number of sources. To reduce the mentioned issues of the sample covariance matrix the framework computes a \ac{ftmr} matrix from the measurements. Then an \ac{evd} decomposes it into eigenvalues and eigenvectors which are utilized to predict the number of sources present and produce a pseudo \ac{music} spectrum, respectively. Trained similarly to \cite{Papageorgiou2020DeepNF}, the localization \ac{nn} input, however, is directly dependent on the selection of noise eigenvectors which makes this method less robust to an incorrect estimation of the number of sources.}

\remove{The works \cite{chakrabarty2019multi, liu2018direction} implement \ac{doa} estimation algorithms in a black-box manner using dense and convolutional \acp{nn}, respectively. The work\add{s \cite{Wu2019DeepCN},} \cite{hammer2021dynamically} trained a \add{\ac{cnn} and} U-Net model\add{, respectively,} to segment a discretized angle grid for identifying multiple \acp{doa}. While these purely \ac{dd} approaches rely less on the accuracy of the assumed data model, they require highly parameterized \acp{nn} that often lack generalization and interpretability, and are computationally intensive.}

\remove{Alternatively, hybrid \ac{mb}/\ac{dd} systems are implemented in \cite{elbir2020deepmusic} and \cite{barthelme2021doa}. Specifically, \cite{elbir2020deepmusic} proposed to estimate the discretized \ac{music} spectrum from the covariance matrix of the measurements through the utilization of multiple convolutional \acp{nn}. However, utilizing the \ac{mb} \ac{music} spectra as training labels inherits the drawbacks of its \ac{mb} counterpart. The work \cite{barthelme2021doa} considers systems with subarray sampling and uses \acp{nn} to obtain a single estimated covariance matrix from incoherent subarray measurements. This \ac{nn}-aided estimate is then utilized for \ac{doa} recovery via the classic \ac{music} algorithm. The method addresses the fundamental dependency of \ac{music} on the estimated covariance matrix; yet again, training using the true covariance matrix as a label does not fully exploit the capability of the \acp{nn}.}

\remove{The works \cite{chakrabarty2017bbdoacnn} and \cite{zhu2019deeplearnbb} both propose a \ac{cnn} architecture for broadband \ac{doa} estimation of one single source. The two approaches, however, utilize very different preprocessing methods and are trained as classification and regression tasks, respectively. The classification \ac{cnn} outputs the posterior probabilities
of the \ac{doa} classes from the phase component of each frequency bin generated by the \ac{fft} of the measurements, thereby limiting itself to a fixed resolution. While the regressor can achieve an arbitrary resolution, it limits itself
to the spatial covariance matrix as input and to the resulting information loss when the measurements are correlated.}

\subsection{MUSIC Algorithm} \label{sec:music}

\begin{figure*}
\centering
\includegraphics[width=0.8\linewidth]{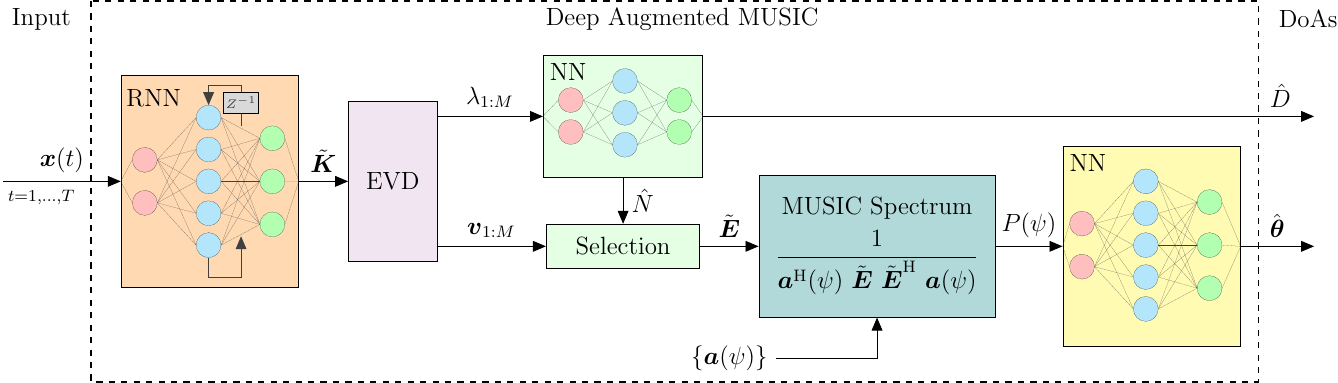}
\caption{Block diagram of the \ac{damusic} algorithm.}
\label{fig:bdAugMUSIC}
\end{figure*}

As our proposed \ac{damusic} algorithm originates from the \ac{music} algorithm, we next present this method in detail. 
\ac{music}, originally proposed by Schmidt in~\cite{schmidt1986music}, considers the narrowband signal model with incoherent sources \eqref{eq:sysmod}, where the signals in $\vect{s}(t)$ are mutually independent. 
Fig.~\ref{fig:bdMUSIC} visualizes a simple outline of the \ac{music} structure as a block diagram. The approach takes the empirical covariance matrix of the received measurements $\mat{X}$, then conducts an \ac{evd}, followed by categorizing the eigenvectors into signal and noise subspaces. The orthogonality between the two subspaces allows the formulation of a spatial spectrum, which contains peaks at \ac{doa} angles.

\subsubsection{Formal Derivation}

With the assumption of the signal and the noise being uncorrelated, the  covariance matrix of $\vect{x}(t)$ is given by
\begin{equation}\label{eq:covariance}
    \mat{K}_{\mat{X}} = 
      \mat{A}(\boldsymbol{\theta}) \mat{K}_{\mat{S}} \mat{A}^\hermit(\boldsymbol{\theta}) + \lambda \mat{K}_{\mat{X}}^0,
\end{equation}
with $\mat{K}_{\mat{S}}$ being the covariance of the incident signals $\vect{s}(t)$. 
The matrix $\mat{A}(\boldsymbol{\theta}) \mat{K}_{\mat{S}} \mat{A}^\hermit(\boldsymbol{\theta})$ is singular and has a rank of less than $M$ when the number of array elements $M$ is strictly larger than the number of signals $D$. Therefore, $\lambda$ is an eigenvalue of $\mat{K}_{\mat{X}}$ (in the metric of $\mat{K}_{\mat{X}}^0$, which takes the form $\mat{K}_{\mat{X}}^0 = \sigma^2 \mathbb{I}$ for \ac{awgn} with variance $\sigma^2$). 
Further, $\mat{A}(\boldsymbol{\theta}) \mat{K}_{\mat{S}} \mat{A}^\hermit(\boldsymbol{\theta})$ has to be non-negative definite, because $\mat{A}(\boldsymbol{\theta})$ has full rank and $\mat{K}_{\mat{S}}$ is positive definite, and consequently, $\lambda$ in \eqref{eq:covariance} is the minimal eigenvalue of $\mat{K}_{\mat{X}}$, denoted $\lambda_{\min}$.
The multiplicity of $\lambda_{\min}$ corresponds to the number of incident wavefronts, and equals $N = M - D$.

\ac{music} builds upon this representation of the covariance of the signals. The algorithm takes as input $T$ snapshots of the waveforms at $M$ array elements, represented as $\mat{X}$ in \eqref{eqn:XMat}, and uses them to  obtain an empirical estimate of $\mat{K}_{\mat{X}}$ via $\hat{\mat{K}}_{\mat{X}} =\frac{1}{T} \mat{X} \mat{X}^\hermit$.
Then, the number of incident signals $D$ is estimated via
\begin{equation}\label{eq:d_estimate}
    \hat D = M - \hat N,
\end{equation}
where $\hat N$ is the estimated multiplicity of the minimal eigenvalue of $\hat{\mat{K}}_{\mat{X}}$. 
The eigenvectors corresponding to the $\hat{N}$ smallest eigenvalues form the noise subspace $\mat{E}_{\hat{N}}$, which is orthogonal to the $D$ dimensional signal subspace spanned by the incident signal mode vectors. 
Consequently, \ac{music} estimates the \acp{doa} by computing the spatial spectrum
\begin{equation}\label{eq:spectrum}
    {P}(\psi) = {1 \over \vect{a}^\hermit(\psi) \mat{E}_{\hat{N}} \mat{E}_{\hat{N}}^\hermit \vect{a}(\psi)},
\end{equation}
and the $\hat D$ dominant peaks of ${P}(\psi)$ are set as the estimated \ac{doa} angles $\hat{\boldsymbol{\theta}}$.

\ac{music} is a popular and highly applicable subspace-based method that is reasonably efficient and statistically consistent~\cite{kay1993fundamentals, krim1996twodecades}. When the signal model is adequately accurate, it can achieve super-resolution and deliver a highly accurate estimate of the number of signal sources present. Nevertheless, the algorithm is sensitive towards the accuracy of the empirical estimate of $\mat{K}_{\mat{X}}$,  and cannot reliably estimate the \ac{doa} angles as well as the number of sources of coherent signals. The reason for this is that highly correlated signals cause zero entries within the covariance matrix and can, therefore, become indistinguishable from noise~\cite{krim1994smooth}. Furthermore, the \ac{music} algorithm is inherently a narrowband approach due to the assumptions imposed on the system model. Nonetheless, it can be extended to broadband, i.e., signal models as in \eqref{eq:bbsysmodfreqVM}, and so we next describe the coherent method to achieve this, which is also adopted in our proposed \ac{damusic}.



\subsection{Coherent Broadband} \label{sec:coherentBB}
The main concept behind coherent broadband \ac{doa} estimation is to transform the different frequency covariance matrices into a single covariance matrix at a focusing frequency. Accordingly, coherent methods find appropriate transformations for each frequency, transform the covariances, and obtain a focused covariance matrix by some form of averaging.
In particular, the \ac{css} method~\cite{wang1985coherent} aims at  combining the spatial signal subspaces to align the signal subspaces associated with the \ac{doa} along all frequency bins. To formulate this, let  $\mat{K}(\omega)$ be the covariance of the frequency domain observations \eqref{eq:bbsysmodfreqVM}, and divide the spectrum into $B$ \acp{ifb} with central frequencies $\{\omega_b\}_{b=1}^B$. 
The focused covariance matrix, which is used as the input covariance for narrowband \ac{doa} recovery, is estimated as 
\begin{equation} \label{eq:focCovMat}
    \mat{K} = \sum_{b=1}^B \alpha_b \ \mat{T}_b \ \mat{K}(\omega_b) \ \mat{T}^\hermit_b,
\end{equation}
where\remove{ $\alpha_b$ is a weighting and} $\mat{T}_b$ is the focusing matrix \add{and frequency bins are prioritized by the weighting $\alpha_b$.} 
The focusing matrices can be determined by attempting to focus the spectral components at some frequency $\omega_r$ and some focusing angles  $\vect{\psi}$  by solving
\begin{equation} \label{eq:focusing}
    \mat{T}_b = \arg \min_{\mat{T}} \|\mat{A}(\omega_r, \vect{\psi}) - \mat{T} \ \mat{A}(\omega_b, \vect{\psi})\|_\forb.
\end{equation}


Coherent techniques have been shown to achieve a better estimation accuracy and a smaller computational complexity as well as a lower resolution threshold than non-coherent methods~\cite{ma2019wideband}. However, the \ac{css} methods require initial values for the focusing matrix $\mat{T}$, the reference frequency $\omega_r$, and the relevant focusing angles $\psi$ to find the focusing matrices with \eqref{eq:focusing}, and are typically sensitive towards these initial values. Additionally, it is not guaranteed that the alignment of the signal and noise subspaces exists to form a viable general covariance matrix without disarranging the noise subspace \cite{yoon2006tops}.

\section{Deep Augmented MUSIC} \label{sec:da-music}
\Ac{damusic} is a hybrid \ac{mb}/\ac{dd} \ac{doa} estimation algorithm derived from the classic \ac{music} algorithm by replacing crucial and model mismatch sensitive elements of the model-based structure with specific \acp{nn}. Fig.~\ref{fig:bdAugMUSIC} outlines the resulting structure of the \ac{damusic} architecture and highlights the remaining similarities to the original \ac{music} algorithm, depicted in Fig.~\ref{fig:bdMUSIC}.

Principally, \ac{damusic} builds upon the understanding that the core challenges associated with the classic \ac{music} algorithm can be tackled by providing a surrogate covariance matrix. In particular, as the \ac{css} method provides a surrogate covariance $\mat{K}$ that transforms a broadband signal model into a narrowband one, a similar approach can be employed to handle coherent sources and array mismatches. Therefore, to improve the categorization of the noise and signal subspaces, the correlation of the received measurements is learned from temporal data by employing a dedicated \ac{rnn}, which is augmented into the overall flow of \ac{music}. 
We next elaborate on the architecture in Subsection~\ref{sec:damArch}, after which we present the training method in Subsection~\ref{sec:damTrain} and provide a discussion in Subsection~\ref{sec:damDisc}.

\subsection{Architecture} \label{sec:damArch} 
\begin{figure}
\centering
\includegraphics[width=0.95\columnwidth]{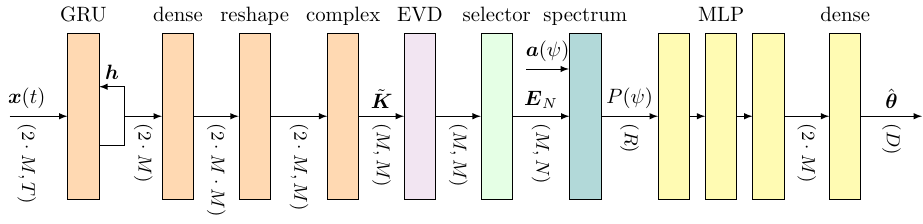}
\caption{Detailed network structure of the \ac{damusic} algorithm.}
\label{fig:nnAugMUSIC}
\end{figure}
The proposed \ac{damusic} algorithm preserves the structure of the \ac{mb} \ac{music} while replacing certain critical aspects with \acp{nn}. Our neural augmentations aim to improve the crucial steps of estimating the noise and signal subspaces from the empirical covariance and the translation of the spatial spectrum  into \acp{doa} via peak finding. By doing so,   \ac{damusic} is not constrained by the additional model assumptions imposed in the derivation of \ac{music}, and can, as we will show, e.g. learn to successfully localize coherent signals.
To present the architecture of \ac{damusic}, we commence with the simple case where the number of sources $D$ is known, and then show how its estimation is incorporated. 
Details of the \acp{nn} used in our experimental study are reported in Section~\ref{sec:numEvals}.

\subsubsection{Known Number of Sources}
Fig.~\ref{fig:nnAugMUSIC} depicts a  detailed outline of the individual elements of the \ac{damusic} architecture. 
The respective output dimensions of the corresponding components are given in the bracket notation. 
First, the input signal $\vect{x}(t)$ is transformed into the pseudo covariance matrix $\Tilde{\mat{K}}$ using a \ac{rnn} implemented through a \ac{gru}. The final state of the \ac{gru} is passed to a dense layer enabling a reshaping to the desired dimension of the pseudo covariance matrix $\Tilde{\mat{K}}$ as well as the subsequent transformation of the complex space. Then, through the continued use of the \ac{evd}, the algorithm categorizes the subspaces using the eigenvectors. 
Inserting the steering vectors $\vect{a}(\psi)$ allows to compute an estimate of the spatial spectrum in \eqref{eq:spectrum}, denoted $P(\psi)$, identically to \ac{mb} \ac{music}, by using the noise eigenvectors selected from  $\Tilde{\mat{K}}$.

Next, \ac{damusic} attains the  \acp{doa} from the spatial spectrum $P(\psi)$ using an additional \ac{nn}, comprised of a \ac{mlp}  of three fully connected dense layers followed by a single dense layer with linear activation. The input to the \ac{mlp} are $R$ samples of $P(\psi)$ taken uniformly in $[0,2\pi)$. 
The output of the \ac{mlp} is the set of estimated \ac{doa} angles $\hat{\boldsymbol{\theta}}$. The benefits of using a \ac{nn}-based peak-finder compared to a model-based one are two-fold. First, learning the translation of the pseudo-spectrum into \acp{doa} from data enables achieving improved resolution compared to conventional peak finding, since $\hat{\theta}_d \in [0, 2\pi)$ instead of being dependent on the number of angles $\psi$ used to evaluate the spectrum. Furthermore, peak finding is generally non-differentiable; thus, replacing it with a \ac{nn} facilitates training  \ac{damusic} end-to-end. The resulting architecture enables the application of gradient-based optimization, by propagating through the \acp{nn} as well as the \ac{evd} operation, as done in \cite{solomon2019deep}. Doing so allows us to jointly tune the noise subspace recovery along with the translation of the \ac{music} spectrum  into \acp{doa} by comparing its estimated \acp{doa} with the true \acp{doa}, as we detail in Subsection~\ref{sec:damTrain}.



\subsubsection{Varying Number of Sources}
Delivering unbiased estimates of the number of signal sources as well as the ability to successfully localize these sources makes  \ac{music}  highly applicable. The above-discussed \ac{damusic} architecture can be extended to operate with a potentially unknown and varying number of signal sources, despite  the determinant nature of the \acp{nn}. This is achieved by addressing three key aspects of the algorithm: the  estimation of the number of sources; the selection process of the noise subspace from the eigenvectors; and the adaptation of the output strategy to overcome the varying number of \ac{doa} angles. 

\begin{figure}
\centering
\includegraphics[width=1.\columnwidth]{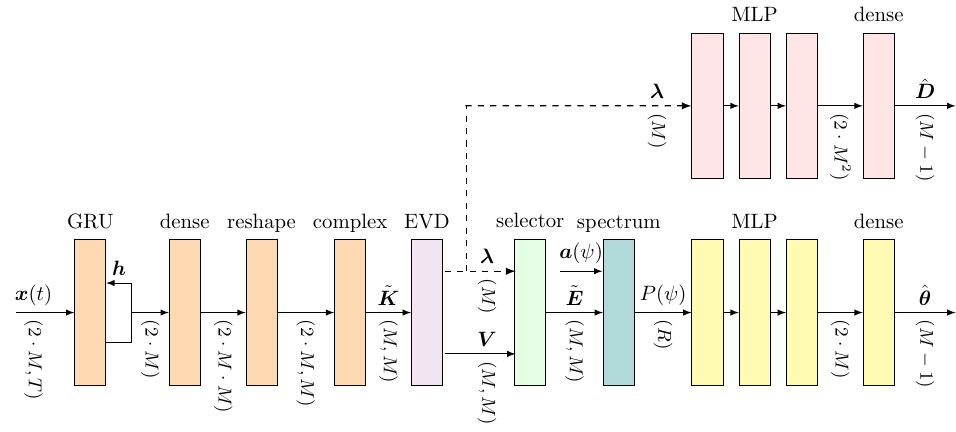}
\caption{\Ac{damusic} algorithm with a separately trained (internal) classifier.}
\label{fig:bdAugMUSICtot}
\end{figure}

\subsubsubsection{Estimating number of sources}
Our design allows \ac{damusic} to learn abstract mappings as pseudo covariance features $\mat{\Tilde{K}}$, which is geared towards end-to-end training and is not restricted to a natural ordering of the model-based covariances. Consequently, instead of estimating the number of sources by inspecting the magnitude of its eigenvalues, we opt a data-driven approach.
We augment the process of estimating the number of sources  with a  classifier, implemented through a \ac{mlp}, as a classification task. Since subspace methods with $M$ inputs can resolve at most $M-1$ signals, the classifier has $M-1$ classes. Fig.~\ref{fig:bdAugMUSICtot} depicts the \ac{damusic} architecture with an added classifier taking the eigenvalues as an input and outputting an estimate of the number of sources.

\begin{figure}
\centering
\includegraphics[width=0.625\columnwidth]{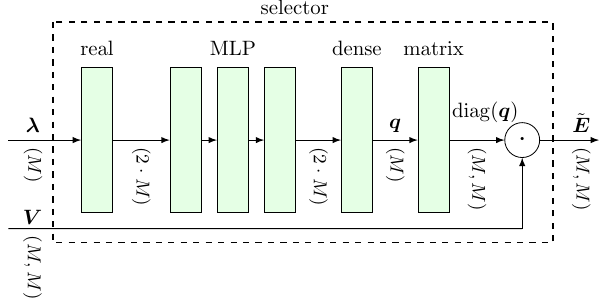}
\caption{Detailed outline of the \ac{damusic} subspace selection augmentation.}
\label{fig:nnAugMUSICsel}
\end{figure}

\subsubsubsection{Computing noise subspace}
The noise subspace selector needs to know the number of sources $D \in \{1, ..., M-1\}$ to classify the eigenvectors into signal and noise subspaces; i.e., by choosing the eigenvectors corresponding to the $N = M - D$ smallest eigenvalues. When the number of sources is not known, we propose to weight each eigenvector by an estimate of it belonging to the noise subspace. We do this by introducing an additional neural augmentation, depicted in Fig.~\ref{fig:nnAugMUSICsel}, which uses a \ac{mlp} to cluster the eigenvalues in a learned fashion. 

In particular, the \ac{mlp} maps the estimated $M$ eigenvalues into a vector~$\vect{q} = \begin{bmatrix}q_1, ..., q_M\end{bmatrix}$ whose entries hold the individual probabilities of choosing the corresponding eigenvector as a noise eigenvector. The selection is performed by computing $\Tilde{\mat{E}} = \mat{V}  \rm{diag}(\vect{q}) = \begin{bmatrix}q_1 \vect{v}_1 \dots q_M \vect{v}_M\end{bmatrix}$, allowing  to learn a suitable noise subspace. 
Note that this setting specializes in the conventional approach of assigning based on the multiplicity of the minimal eigenvalues; in the conventional approach, the entries of $\vect{q}$ are either ones or zeros and $\Tilde{\mat{E}}$ coincides with the corresponding subspace. Consequently, the proposed approach provides additional flexibility in selecting the noise subspace and facilitates coping with settings where distinguishing between the eigenvectors is challenging due, for example, to low \acp{snr}.  

\subsubsubsection{Outputting varying number of \acp{doa}}
The most common strategy to overcome the dynamic output dimensions of \acp{nn} is to scale the output dimension to the maximum occurring value. In our case, since \ac{music} cannot resolve more than $M-1$ sources,  the dimension of the last dense layer of \ac{damusic} is set to $M-1$. The only additional modification required with this strategy compared with known $D$ is the slight alteration to the loss function discussed in Subsection~\ref{sec:damTrain} below. 
An advantage of this strategy is that the approach allows to extract up to $M-1$ \ac{doa}, where $\theta_1$ is most likely a true \ac{doa} angle, $\theta_2$ has a slightly lower probability to be a \ac{doa} angle, etc. The final estimation is thus carried out by first taking $\hat{D}$ from the module that estimates the number of sources, and then using the first $\hat{D}$ outputs as the recovered \acp{doa}.

\subsection{Training Procedure} \label{sec:damTrain}
\ac{damusic} is trained end-to-end in a supervised setting as a multiple regression problem. As detailed in Subsection~\ref{sec:problem}, the training set is comprised of $U$ tuples of sequences of measurements and their corresponding \acp{doa}; i.e., the $u^{\text{th}}$ tuples includes the $T_u$ measurements $\mat{X}_{u}$ and their corresponding $D_u$ \ac{doa} angles $\boldsymbol{\theta}_u$. We first describe how this data is used for training when $D$ is known and fixed; i.e., $D_u \equiv D$. This acts as a preliminary step for discussing how the training procedure is carried out in the general case where the model does not have knowledge $D$.

\subsubsection{Known Number of Sources}
Given a sequence of measurements $\mat{X}$ as input, the model predicts the estimated \ac{doa} angles $\hat{\boldsymbol{\theta}}$ that are compared to the true \ac{doa} angles $\boldsymbol{\theta}$. Gradient-based optimization is possible because every element of the architecture is differentiable, allowing backpropagation through the complete structure. 
Derived from the \ac{rmspe}~\cite{routtenberg2011bayesian, routtenberg2013non-bayesian} the following loss function additionally compares all permutations of the predicted angles with the true angles to capture all possible assignments of the estimated \ac{doa} to the true \ac{doa}. Thereby the minimal permutation \ac{rmspe} includes the permutation invariance of the \acp{doa} and is obtained as:
\begin{equation}
\label{eq:rmspe}
    \text{RMSPE}(\boldsymbol{\theta}, \boldsymbol{\hat \theta}) = \min_{\mat{P} \in \perm{P}_D}\left(\frac{1}{D} \left\|\text{mod}_{\beta}(\boldsymbol{\theta} - \mat{P}\boldsymbol{\hat \theta})\right\|^2\right)^\frac{1}{2},
\end{equation}
where $\perm{P}_D$ is the set of all $D\times D$ permutations and $\text{mod}_{\beta}$ denotes the element-wise modulus operation regarding the angle range of interest, e.g. $\beta = \pi$ for $\psi \in [-\pi/2, \pi/2)$ or $\beta = 2 \pi$ for $\psi \in [0, 2 \pi)$.

\subsubsection{Unknown Number of Sources}
As discussed in the previous subsection, \ac{damusic} is designed to resolve a varying and unknown number of sources by introducing an additional \ac{nn} classifier for the number of sources. To formulate the training procedure of the overall system, we use $\vect{w}_c$ to denote the trainable parameters of the classifier, while $\vect{w}_{d}$ represents the parameters of the remaining trainable modules of \ac{damusic} (covariance estimator, peak finder, and subspace selector). 
The loss used to train $\vect{w}_{d}$ is the \ac{rmspe} loss of \eqref{eq:rmspe}, which is altered during training to account for varying sources by computing,
\begin{equation}
\label{eqn:rmspe2}
    \mathcal{L}_{\rm RMSPE}\big(\boldsymbol{\theta}_u, \boldsymbol{\hat{\theta}}(\mat{X}_u)\big) = \text{RMSPE}\big(\theta_{1:D_u}, \hat{\theta}_{1:D_u}(\mat{X}_u)\big).
\end{equation}
In \eqref{eqn:rmspe2}, $\boldsymbol{\hat{\theta}}(\mat{X}_u)$ denotes the $M-1$ outputs of \ac{damusic} applied to $\mat{X}_u$, while $\theta_{1:D_u} = (\theta_{1}, ..., \theta_{D_u})$. This means that only the first $D_u$ angles of the estimated \ac{doa} $\boldsymbol{\hat{\theta}}$ are compared with the true \ac{doa} $\boldsymbol{\theta}$ while the remaining angles of $\boldsymbol{\hat{\theta}}$ are completely ignored.

The separate classifier is trained using the categorical cross-entropy of the classes as a loss function ensuring optimal training. In particular, letting $\boldsymbol{\lambda}_u$ be the input to the \ac{mlp} classifier when applying \ac{damusic} to $\mat{X}_u$ and letting $\boldsymbol{\hat{D}}(\vect{\lambda}_u)$ be the softmax output of the \ac{mlp} applied to $\boldsymbol{\lambda}_u$ (with ${\hat{D}}_i(\vect{\lambda}_u)$ being its $i$th entry), the loss used for training $\vect{w}_{c}$ is given by
\begin{equation}
\label{eqn:ce}
    \mathcal{L}_{\rm CE}\big(D_u, \boldsymbol{\hat{D}}(\boldsymbol{\lambda}_u)\big) = -\log{\hat{D}}_{D_u}(\vect{\lambda}_u).
\end{equation} 
It is noted that since \eqref{eqn:ce} is used for training $\vect{w}_{c}$, then one should  block the gradient during backpropagation from passing from the classifier to the \ac{evd}, as indicated by the dashed connections in Fig.~\ref{fig:bdAugMUSICtot}. This ensures that the \ac{mlp} learns from the eigenvalues themselves and not by influencing and disrupting the \ac{gru}. Further, the regressor is completely independent of the classifier, which allows \ac{damusic} to operate with different classifiers if needed or with any other desired scheme delivering an estimate of $D$. The resulting training procedure (employing  mini-batch gradient descent with ADAM \cite{kingma2014adam}) is summarized as Algorithm~\ref{alg:Training}.

\RestyleAlgo{ruled}
    \begin{algorithm}
    {\fontsize{9pt}{9pt}\selectfont
    \caption{Training \ac{damusic} 
    }\label{alg:Training} 
    \KwData{Data set $\{(\mat{X}_u, 
 \boldsymbol{\theta}_u)\}_{u=1}^{U}$, learning rate $\mu=0.001$, decay rates $b_1=0.9, b_2=0.999$, $\epsilon=10^{-8}$}
    Initialize weights $\vect{w}_{d}, \vect{w}_{c}$\;
    Initialize moment vectors $\boldsymbol{\nu}_{d}, \boldsymbol{\upsilon}_{d}, \boldsymbol{\nu}_{c}, \boldsymbol{\upsilon}_{c}$\;
   \For{${\rm epoch}=1,2,\ldots$}{
        \For{$\rm each$ $\rm batch$}{
            Apply \ac{damusic} to $\{\mat{X}_u\}$ for $u\in {\rm batch}$\;
            Compute gradients $\vect{g}_{d}$ via $\vect{g}_{d} \leftarrow \nabla_{\vect{w}_{d}}\sum\limits_{u \in {\rm batch}}    \mathcal{L}_{\rm RMSPE}(\boldsymbol{\theta}_u, \boldsymbol{\hat{\theta}}(\mat{X}_u))$\;
            Compute gradients $\vect{g}_{c}$ via
            $\vect{g}_{c} \leftarrow \nabla_{\vect{w}_{c}}\sum\limits_{u \in {\rm batch}}    \mathcal{L}_{\rm CE}(D_u, \boldsymbol{\hat{D}}(\boldsymbol{\lambda}_u)) $\;
            Update biased first moment via\\
             \hspace{2mm}$\boldsymbol{\nu}_d \leftarrow b_1 \boldsymbol{\nu}_d + (1-b_1) \vect{g}_{d}$\;
             \hspace{2mm}$\boldsymbol{\nu}_c \leftarrow b_1 \boldsymbol{\nu}_c + (1-b_1) \vect{g}_{c}$\;
            Update biased second raw moment via\\
            \hspace{2mm}$\boldsymbol{\upsilon}_d \leftarrow b_2 \boldsymbol{\nu}_d + (1-b_2) \vect{g}_{d}^2$\;
             \hspace{2mm}$\boldsymbol{\upsilon}_c \leftarrow b_2 \boldsymbol{\nu}_c + (1-b_2) \vect{g}_{c}^2$\;
            Compute bias corrected moments $\hat{\boldsymbol{\nu}}_d, \hat{\boldsymbol{\nu}}_c, \hat{\boldsymbol{\upsilon}}_d, \hat{\boldsymbol{\upsilon}}_c$\;
            Update $\vect{w}_{d}$ via $\vect{w}_{d} \leftarrow \vect{w}_{d} - \mu \ \hat{\boldsymbol{\nu}}_d / (\sqrt{\hat{\boldsymbol{\upsilon}}_d} + \epsilon)$\;
            Update $\vect{w}_{c}$ via $\vect{w}_{c} \leftarrow \vect{w}_{c} - \mu \ \hat{\boldsymbol{\nu}}_c / (\sqrt{\hat{\boldsymbol{\upsilon}}_c} + \epsilon)$\;
            }
        }}
    \end{algorithm}


\subsection{Discussion} \label{sec:damDisc}
The design of the  architecture of \ac{damusic} is derived from the model-based \ac{music} structure. This allows for exploiting the successful aspects of the algorithm while improving certain critical elements and alleviating important drawbacks. Replacing the empirical covariance estimation with a \ac{rnn} is the key neural augmentation of \ac{damusic}, enabling the system to learn the pseudo covariance from the measurements themselves such that the resulting surrogate model facilitates subspace-based \ac{doa} recovery. Thereby, the performance of \ac{damusic}, for example, is not affected by coherent signals and other related issues discussed in Subsection~\ref{sec:music}. Furthermore, learning end-to-end allows \ac{damusic} to operate with broadband signals, as it effectively learns to produce a focused  pseudo covariance, similarly to the \ac{css} methods discussed in Subsection~\ref{sec:literature}. It is noted that our augmentation approach depends on the array geometry, as the steering vectors $\vect{a}(\cdot)$ are used for computing the \ac{music} spectrum. Nonetheless, as we numerically demonstrate in Section~\ref{sec:numEvals}, \ac{damusic} learns to overcome mismatches in the array geometry from the data without any alterations or renewed training.

Specifically, the \ac{rnn} utilized by \ac{damusic} is able to learn an appropriate focusing matrix while concurrently correlating the measurements comparably to~\eqref{eq:focCovMat}. 
The usage of a \ac{rnn} applied to the time-domain signal and only passing the last state to the next layer allows \ac{damusic} to operate with different signal durations and possibly cope in real-time with dynamic variations in the \acp{doa}, though investigation of the latter is left for future work.
Our experiments indicate that a deeper \ac{rnn} aids the correlation aspect while a wider \ac{rnn} allows a more complicated mapping with this correlation. 

\begin{table}
    \centering
    \caption{Simulation parameters.}
    \label{tab:simParams}
    \begin{tabular}{|l|c|c|} 
    \hline
    Parameter Description & Notation & Default Value\\ 
    \hline
    \hline
    Array geometry & & \ac{ula} \\
    Number of array elements & $M$ & 8 \\
    Element spacing & $\Delta m$ & $\ell_{\min}/2$ \\
    \Ac{snr} & & 10 dB \\ 
    Snapshots & $T$ & 200 \\
    Grid points of continuum & $R$ & 360 \\
    Min. frequency & $f_{\min}$ & $0 \Hz$ \\
    Max. frequency & $f_{\max}$ & $999 \Hz$ \\
    Sampling frequency & $f_s$ & $2 (f_{\max} + 1)$ \\
    Time length & $T_{samp}$ & 1 s \\
    \Ac{fft} points & $N_f$ & $T_{samp} \cdot f_s$ \\
    \hline
    \end{tabular}
\end{table}

Another important component of the architecture of \ac{damusic} is its incorporation of the \ac{evd} as a means for division into subspaces. Though computationally expensive, the internal \ac{evd} allows \ac{damusic} to not only classify signal and noise subspaces with the eigenvalues, but also significantly simplifies the estimation of the number of signal sources present. 

To enable training end-to-end from the errors in \eqref{eq:rmspe}, a \ac{nn}-based peak finder is used in form of a \ac{mlp}. Model-based peak-finding is generally non-differentiable; therefore, replacing it with a \ac{nn} enables gradient-based optimization through the entire \ac{damusic} structure. Furthermore, improved resolution can be achieved by extracting the \ac{doa} from the spatial spectrum in a learned manner (i.e., $\hat{\theta}_d \in [0, 2\pi)$ and it is not dependent on the number of angles $\psi$ used to evaluate the spectrum $P(\psi)$).


\section{Numerical Evaluations} \label{sec:numEvals}
In this section, we present our numerical evaluations of the proposed \ac{damusic} algorithm\footnote{The source code used in our experiments can be found at \url{https://github.com/DA-MUSIC/TVT23}.} Our experimental study is comprised of evaluations in a narrowband synthetic setting (Subsection~\ref{sec:synthNb}); a broad synthetic setup (Subsection~\ref{sec:synthBb}); and experiments with real-world data corresponding to azimuth estimation in seismic arrays (Subsection~\ref{sec:non-synth}). 

\subsection{Synthetic Narrowband} \label{sec:synthNb}

The numerical evaluations of synthetic data presented below are obtained by simulating the measurements $\vect{x}(t)$ according to the narrowband system model~\eqref{eq:sysmod}. In particular, we simulate a \ac{ula} with $M=8$ array elements that measure impinging waveforms originating from the \ac{doa} angles $\boldsymbol{\theta} = [\theta_1, \dots, \theta_D]$, which are separately drawn from the uniform distribution $\mathcal{U}(-\pi/2, \pi/2)$. The signals $\vect{s}(t) = [s_1(t) \dots s_D(t)]^\transpose$ are each drawn randomly from the complex Gaussian distribution $\mathcal{CN}(0, 1)$ for all $t$ modeling random amplitudes and phases. The noises measured at the $M$ array elements $\vect{v}(t) = [v_1(t) \dots v_M(t)]^\transpose$ are also drawn from $\mathcal{CN}(0, 1)$ for all $t$, followed by appropriate scaling to meet the constant \ac{snr}. In the coherent cases, all signals have identical amplitudes and phases, and when not stated otherwise, the simulation parameters are set according to Table~\ref{tab:simParams}.

\subsubsection{Known Number of Sources} \label{sec:resKnownNumS}
We first evaluate the \ac{rmspe} in [rad] achieved when the number of sources $D$ is known in the synthetic narrowband scenario described above. The results, reported in Table~\ref{tab:results}, compare the performance of the following \ac{doa} estimators for a different number of sources $D$: 
\begin{itemize}
    \item The \ac{damusic} architecture is implemented according to Fig.~\ref{fig:bdAugMUSIC} and  trained separately for each case $D$.
    \item The classic \ac{mb} MUSIC algorithm, implemented as described in Section~\ref{sec:music}, utilizes the external knowledge of $D$.
    \add{\item The \ac{cnn} architecture of \cite{Papageorgiou2020DeepNF}, with a reduced stride in the first layer to account for $M=8$, and an increased grid-size for the last layer to achieve $R=360$.}
    \item The \ac{dd} DeepMUSIC proposed in \cite{elbir2020deepmusic}, while incorporating minor alterations that were necessary to accommodate for the difference in the setup, 
    includes tuning of individual hyperparameters to assure successful training of the \acp{cnn}.
\end{itemize}
The above algorithms are compared with a random guessing of the \ac{doa} angles. 
The results in Table~\ref{tab:results} show that the proposed \ac{damusic} notably outperforms all considered benchmarks, notably surpassing the \ac{mb} MUSIC algorithm not only for coherent sources, but also for non-coherent ones, which is the scenario for which the \ac{mb}  algorithm is designed. 

\begin{table}[b]
    \centering
    \caption{\Ac{rmspe} of different \ac{doa} estimation algorithms with constant and known $D$ for $T=200$ snapshots.}
    \label{tab:results}
    \begin{tabular}{ | c | c | c | c | c | c | } 
        \hline
        \thead{RMSPE\\{[rad]}} & \thead{DA-MUSIC} & \thead{CNN} & \thead{Deep-\\MUSIC} & \thead{Classic\\MUSIC} & 
        \thead{Random} \\
        \hline
        \hline
        \multicolumn{6}{| c |}{non-coherent} \\
        \hline

        $D=2$ & \textbf{0.0117} & 0.0237 & 0.0329 & 0.0336 & 
        0.6809 \\
        \hline
        $D=3$ & \textbf{0.0315} & 0.0464 & 0.1600 & 0.0841 & 
        0.6034 \\
        \hline
        $D=4$ & \textbf{0.0563} & 0.0841 & 0.2656 & 0.1459 & 
        0.5421 \\
        \hline
        $D=5$ & \textbf{0.0751} & 0.1319 & 0.2701 & 0.2008 & 
        0.4963 \\
        \hline
        \hline
        \multicolumn{6}{| c |}{coherent} \\
        \hline
        $D=2$ & \textbf{0.0140} & 0.0274 & 0.5781 & 0.2350 & 
        0.6854 \\
        \hline
        $D=3$ & \textbf{0.0407} & 0.0528 & 0.4023 & 0.4819 & 
        0.6060 \\
        \hline
        $D=4$ & \textbf{0.0519} & 0.0859 & 0.2915 & 0.4522 & 
        0.5407 \\
        \hline
        $D=5$ & \textbf{0.0658} & 0.1254 & 0.3054 & 0.4401 & 
        0.5000 \\
        \hline
    \end{tabular}
\end{table}



\begin{figure}
\centering
\includegraphics[width=1.0\columnwidth]{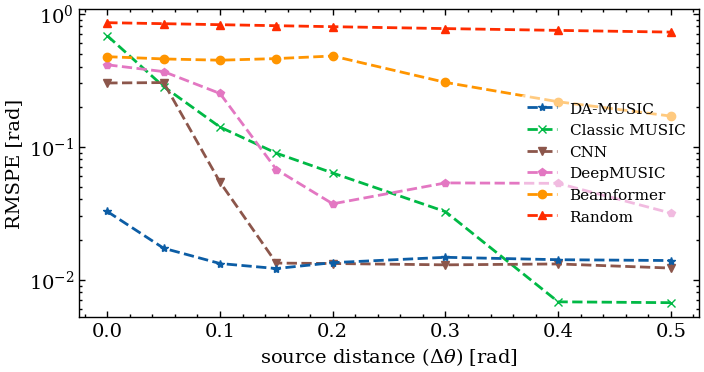}
\caption{\ac{doa} estimation of $D=2$ closely spaced sources.}
\label{fig:resolutionCap}
\end{figure}

\begin{figure}
\centering
\includegraphics[width=1.0\columnwidth]{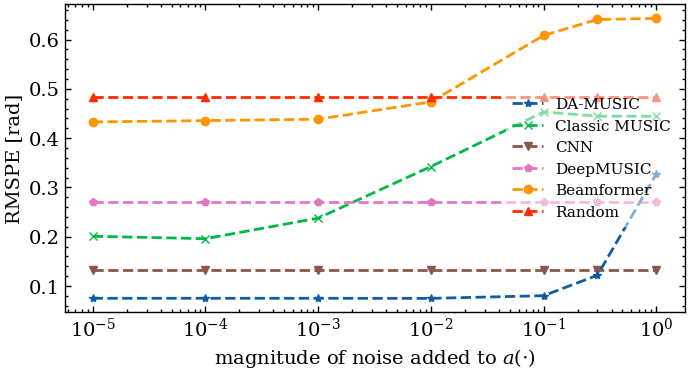}
\caption{DoA estimation with mismatch in the array geometry for $D=5$ sources.}
\label{fig:mismatch}
\end{figure}

\begin{figure}[t]
\centering
\includegraphics[width=1.0\columnwidth]{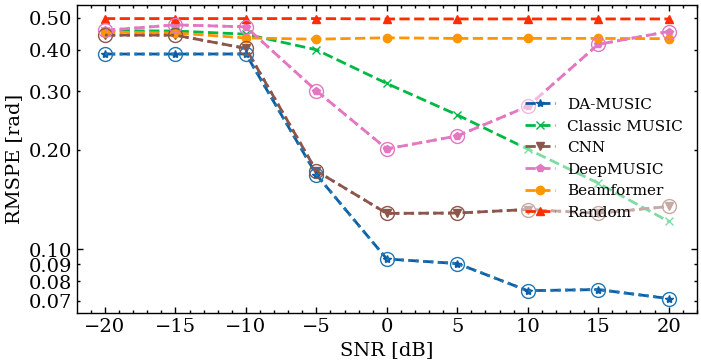}
\caption{DoA estimation of $D = 5$ signals with different SNRs.}
\label{fig:errVsnr}
\end{figure}

\begin{figure}[t]
\centering
\includegraphics[width=1.0\columnwidth]{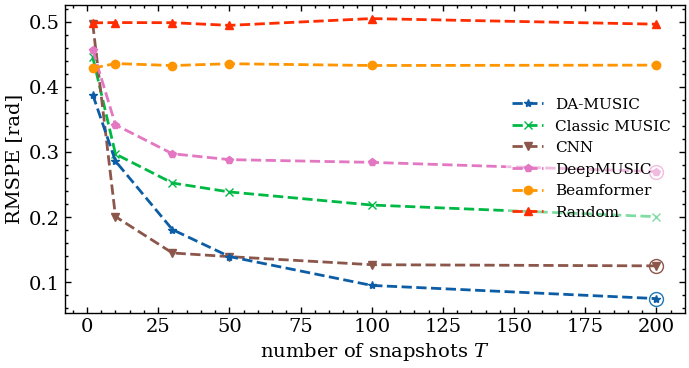}
\caption{DoA estimation of $D = 5$ signals with different number of snapshots $T$.}
\label{fig:errVsnaps}
\end{figure}

To compare the resolution  of the algorithms, Fig.~\ref{fig:resolutionCap} shows the \ac{rmspe} for localizing $D=2$ non-coherent signals, which are located  close together at a $\Delta \theta$ distance from each other. The \ac{mb} \ac{music} algorithm is shown to collapse when the angular difference approaches $\Delta\theta \approx 0.1$ radians, while \ac{damusic} demonstrates a constant low error for all $\Delta\theta$, indicating its improved resolution.

\begin{table}[b]
    \centering
    \caption{\Ac{rmspe} of different \ac{doa} estimation algorithms with varying and unknown $D$ for $T=200$ snapshots.}
    \label{tab:resultsUnD}
    \begin{tabular}{ | c | c | c | c | c | c |} 
        \hline
        \thead{RMSPE\\{[rad]}} & \thead{DA-MUSIC} & \thead{CNN} & \thead{Classic\\MUSIC} & \thead{Beamformer} & \thead{Random} \\
        \hline
        \hline
        \multicolumn{6}{| c |}{non-coherent} \\
        \hline

        $D=2$ & 0.0430 & \textbf{0.0403} & 0.0428 & 0.1900 & 0.8318 \\
        \hline
        $D=3$ & \textbf{0.0705} & 0.0842 & 0.0917 & 0.2906 & 0.6981 \\
        \hline
        $D=4$ & \textbf{0.0894} & 0.1463 & 0.1489 & 0.3757 & 0.6021 \\
        \hline
        $D=5$ & \textbf{0.1222} & 0.1847 & 0.1856 & 0.4029 & 0.5357 \\
        \hline
        \hline
        \multicolumn{6}{| c |}{coherent} \\
        \hline

        $D=2$ & 0.0383 & \textbf{0.0376} & 0.6139 & 0.1670 & 0.8243 \\
        \hline
        $D=3$ & \textbf{0.0688} & 0.0729 & 0.5743 & 0.2843 & 0.6962 \\
        \hline
        $D=4$ & \textbf{0.0869} & 0.1181 & 0.5258 & 0.3883 & 0.6051 \\
        \hline
        $D=5$ & \textbf{0.1046} & 0.1652 & 0.4744 & 0.4179 & 0.5435 \\
        \hline
    \end{tabular}
\end{table}

Next, we evaluate \ac{damusic} in the presence of a mismatch in the array geometry. 
Fig.~\ref{fig:mismatch} depicts the \ac{rmspe} achieved when each element of the steering vector $\vect{a}(\cdot)$ is corrupted with zero-mean Gaussian noise, leading to a mismatch from the values used to compute the spatial spectrum. Indicating improved robustness, \ac{damusic} is shown to overcome such mismatches in the array geometry from the data. \add{The \ac{cnn}, deepMUSIC, and the Random algorithm remain unaffected as they are independent of $\vect{a}(\cdot)$.}

\remove{We conclude the evaluation of \ac{damusic} with a known number of sources by considering the case in which the \ac{snr} is not known during training, and is thus trained using data generated from a mixture of different \ac{snr} levels in the range of $0-20$ dB.}
Fig.\ref{fig:errVsnr} depicts the performances differences when localizing $D = 5$ non-coherent signals with  \add{$T=200$ snapshots available for}\remove{various} different \ac{snr}\add{levels in the range of $[-20, 20]$ dB}. \Ac{damusic} shows a constant low \ac{snr}\remove{even for low}\add{for positive dB} settings and without any fluctuations which slowly decreases with increasing \ac{snr}. 

\add{We conclude the evaluation of \ac{damusic} with a known number of sources by considering the case in which the number of available snapshots $T$ is varied. Fig. \ref{fig:errVsnaps} depicts the performance degradation of the estimators with fewer snapshots available. The \ac{dd} estimators are only trained for the case $T=200$ as indicated by the circle around the $T=200$ marker, yet manage to operate with shorter sequences during inference due to the recurrent unit or by taking the covariance matrix as input.}

\subsubsection{Unknown and Varying Number of Sources}
\begin{figure}[t]
\centering
\includegraphics[width=1.0\columnwidth]{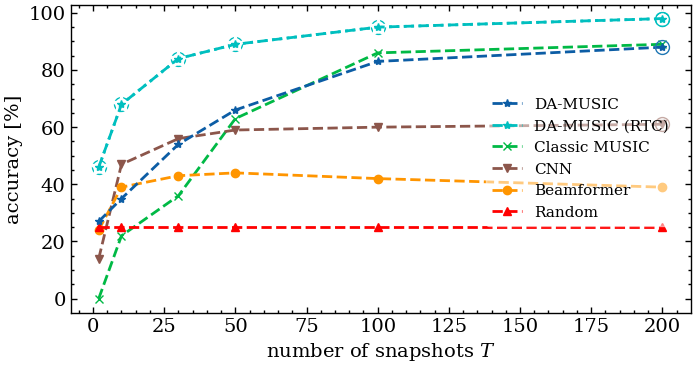}
\caption{Accuracy of estimating $\hat{D}$ for various $T$ with ${D \in \{2, ..., 5\}}$ non-coherent signals.}
\label{fig:varDaccVsnaps}
\end{figure}
\begin{figure}[t]
\centering
\includegraphics[width=1.0\columnwidth]{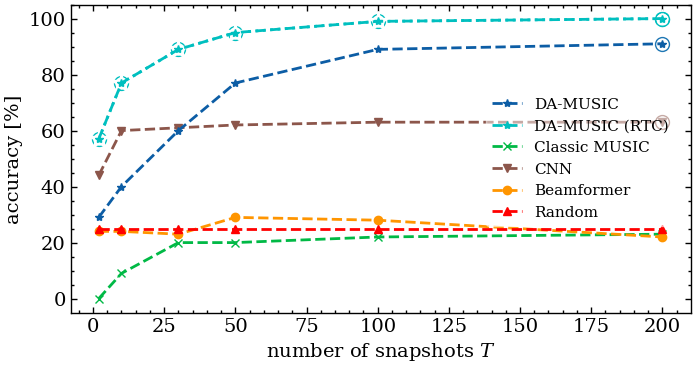}
\caption{Accuracy of estimating $\hat{D}$ for various $T$ with ${D \in \{2, ..., 5\}}$ coherent signals.}
\label{fig:varDaccVsnapsCo}
\end{figure}
\begin{figure}[t]
\centering
\includegraphics[width=1.0\columnwidth]{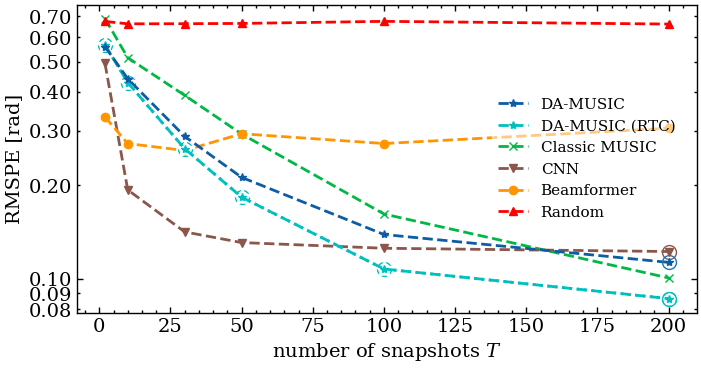}
\caption{RMSPE for varying and unknown number of ${D \in \{2, ..., 5\}}$ non-coherent signals.}
\label{fig:varDerrVsnaps}
\end{figure}
\begin{figure}[t]
\centering
\includegraphics[width=1.0\columnwidth]{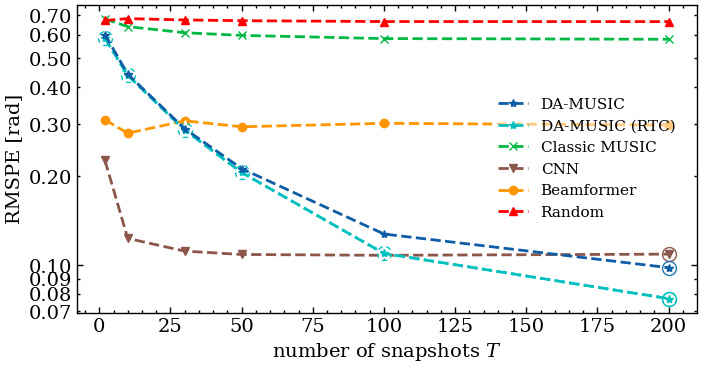}
\caption{RMSPE for varying and unknown number of ${D \in \{2, ..., 5\}}$ coherent signals.}
\label{fig:varDerrVsnapsCo}
\end{figure}
Table~\ref{tab:resultsUnD} shows the results obtained in the exact same narrowband scenario introduced in Section~\ref{sec:synthNb} above but with an unknown and varying number of sources $D \in \{2, ..., 5\}$. Here, during inference, the \ac{doa} estimation algorithms do not have any knowledge of the varying number of sources present. To compute the \ac{rmspe} in such settings (i.e., if the \ac{doa} estimators output the wrong number of sources $\hat{D}$), the \ac{doa} are either truncated (least dominant peaks for the \ac{mb} algorithms \add{and the \ac{cnn}} and highest indexed \ac{doa} angles for \ac{damusic}) or padded with random \ac{doa} angels until $|\hat{\boldsymbol{\theta}}| = |\boldsymbol{\theta}| = D$. We compare the following \ac{doa} estimators:
\begin{itemize}
    \item The \ac{damusic} architecture is implemented according to Fig.~\ref{fig:bdAugMUSICtot} with the classifier which predicts the number of sources being trained along with the overall \ac{doa} estimation method. 
    \item The \ac{damusic} (RTC) variation is also implemented according to Fig.~\ref{fig:bdAugMUSICtot}, yet with a retroactively trained classifier (RTC), i.e., we first train \ac{damusic} for a specific scenario, and then we fix the \ac{doa} estimator modules and train only the classification network for alternate scenarios. 
    \item The Classic MUSIC algorithm is implemented as before but determines the number of sources by estimating the multiplicity of the smallest eigenvalue utilizing a pre-determined threshold.
    \add{\item The \ac{cnn} of \cite{Papageorgiou2020DeepNF} as introduced above.}
    \item A conventional beamformer~\cite{bartlett1948beamformer}, utilizing a peak-finder to estimate the number of sources by determining the number of dominant peaks.
    \item The Random algorithm corresponds to the base performance when choosing \ac{doa} angles at random. 
\end{itemize}

Figs. \ref{fig:varDaccVsnaps} and \ref{fig:varDaccVsnapsCo} show the accuracy in identifying the number of sources versus the number of snapshots $T$ obtained by the mentioned algorithms during the estimation of $\hat{D}$ for non-coherent and coherent signals, respectively. \remove{\ac{damusic} is only trained for the case $T=200$ as indicated by the circle around the $T=200$ marker, yet manages to operate with shorter sequences during inference due to the recurrent unit.}\add{Again, the \ac{dd} estimators are only trained for the case $T=200$ as indicated by the circle.} Unfortunately, the performance of the internal classifier of \ac{damusic} is dependent on the number of snapshots, and to be able to maintain a more constant accuracy it must be trained for each case. Consequently, \Ac{damusic} (RTC), which trains its classifier retroactively, requires separate training for each number of snapshots, yet manages to achieve the most accurate predictions.

The corresponding performances in localizing the varying number of sources are depicted in Figs. \ref{fig:varDerrVsnaps} and \ref{fig:varDerrVsnapsCo} for non-coherent and coherent signals, respectively. The shown \ac{rmspe} is an average over all considered $D \in \{2, ..., 5\}$. The fundamental limitation of the \ac{mb} \ac{music} structure to estimate the number of signal sources for coherent signals also severely impacts the localization abilities of the algorithm. \remove{The \ac{damusic} algorithm outperforms the classic \ac{music} approach in non-coherent and coherent cases.}

\subsection{Synthetic Broadband} \label{sec:synthBb}
\begin{table}
    \centering
    \caption{Simulation parameters of the broadband scenarios.}
    \label{tab:bbSimParams}
    \begin{tabular}{|c|l|c|c|} 
    \hline
     & Parameter Description & Notation & Value\\ 
    \hline
    \hline
    Scenario 1 & Modulation frequency & $f_{c, d}$ & $0 - 999 \Hz$ \\
    \hline
    \hline
    \multirow{2}{*}{Scenario 2} & Number of subcarriers & $K$ & 1000 \\
    & Signal bandwidth & $\Delta f_d$ & $1000 \Hz$  \\
    \hline
    \hline
    \multirow{3}{*}{Scenario 3} & Modulation frequency & $f_{c, d}$ & $0-899 \Hz$ \\
    & Number of subcarriers & $K$ & 10 \\
    & Signal bandwidth & $\Delta f_d$ & $100 \Hz$  \\
    \hline
    \end{tabular}
\end{table}
We proceed to evaluate \ac{damusic} in a broadband setting.
The previously introduced synthetic environment requires certain alterations and assumptions to account for broadband signals. The sensor elements of the receiving array are adequately spaced, and the element spacing is therefore assumed to be
\begin{equation}
    \Delta m = \frac{\ell_{\min}}{2} = \frac{1}{2} \ \frac{c}{f_{\max}},
\end{equation}
where $\ell_{\min}$ is the minimal wavelength corresponding to the maximal occurring  frequency $f_{\max}$ and the frequency spectrum of interest is considered to be within $[f_{\min}, f_{\max}]$. The measurements are simulated utilizing the broadband system model~\eqref{eq:bbsysmodfreqVM}, where the elements of $\vect{\ft{S}}(\omega)$ and $\vect{\ft{W}}(\omega)$ are the $N_f$-point \acp{fft} of the elements of $\vect{s}(t)$ and $\vect{w}(t)$. The parameter values of the different broadband scenarios are specified in Table~\ref{tab:bbSimParams} if not specified otherwise. 
We simulate the following \ac{doa} estimators:
\begin{itemize}
    \item The \ac{damusic} architecture is implemented according to Fig.~\ref{fig:bdAugMUSIC}, but the \ac{gru} parameters are scaled (having 10 times more parameters available) to enable optimal learning despite the more complex broadband scenarios.
    \item The classic MUSIC algorithm is implemented in its narrowband format as described in Section~\ref{sec:music} and utilizes steering vectors calibrated to the exact array element spacing using $\ell_{\min} / 2$.
    \item Broadband \ac{music} corresponds to an incoherent broadband extension of \ac{music} \cite{yoon2006doaest} and is implemented using $10 \Hz$ per \ac{ifb}; i.e., $|\omega_{b}-\omega_{b-1}| = 10 \Hz$ for all ${b \in \{1, ..., B\}}$.
    \item The \ac{doa} estimators are again compared to choosing \ac{doa} angles at random.
\end{itemize}
\begin{figure}
\centering
\includegraphics[width=1.0\columnwidth]{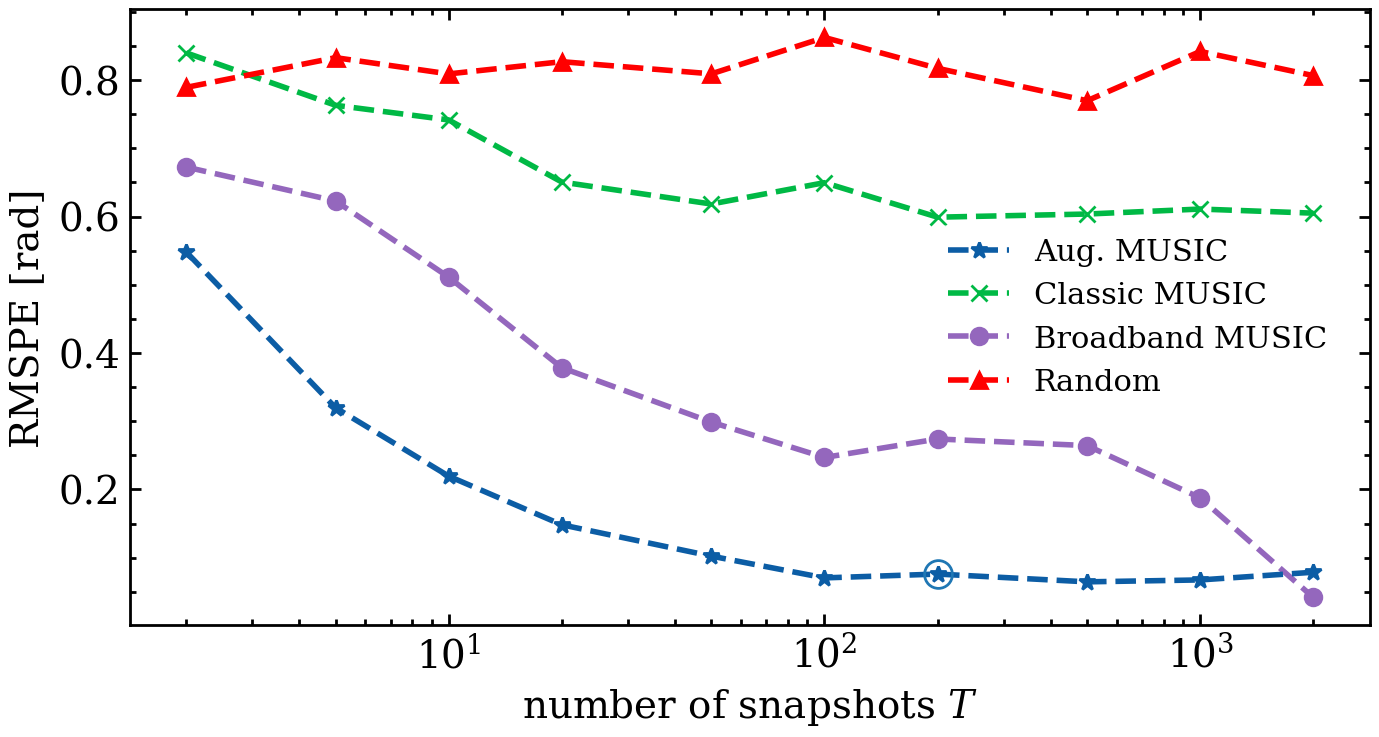}
\caption{RMSPE for known number of $D=2$ signals from Broadband Scenario 1.}
\label{fig:errVsnapsS1}
\vspace{2mm}
\centering
\includegraphics[width=1.0\columnwidth]{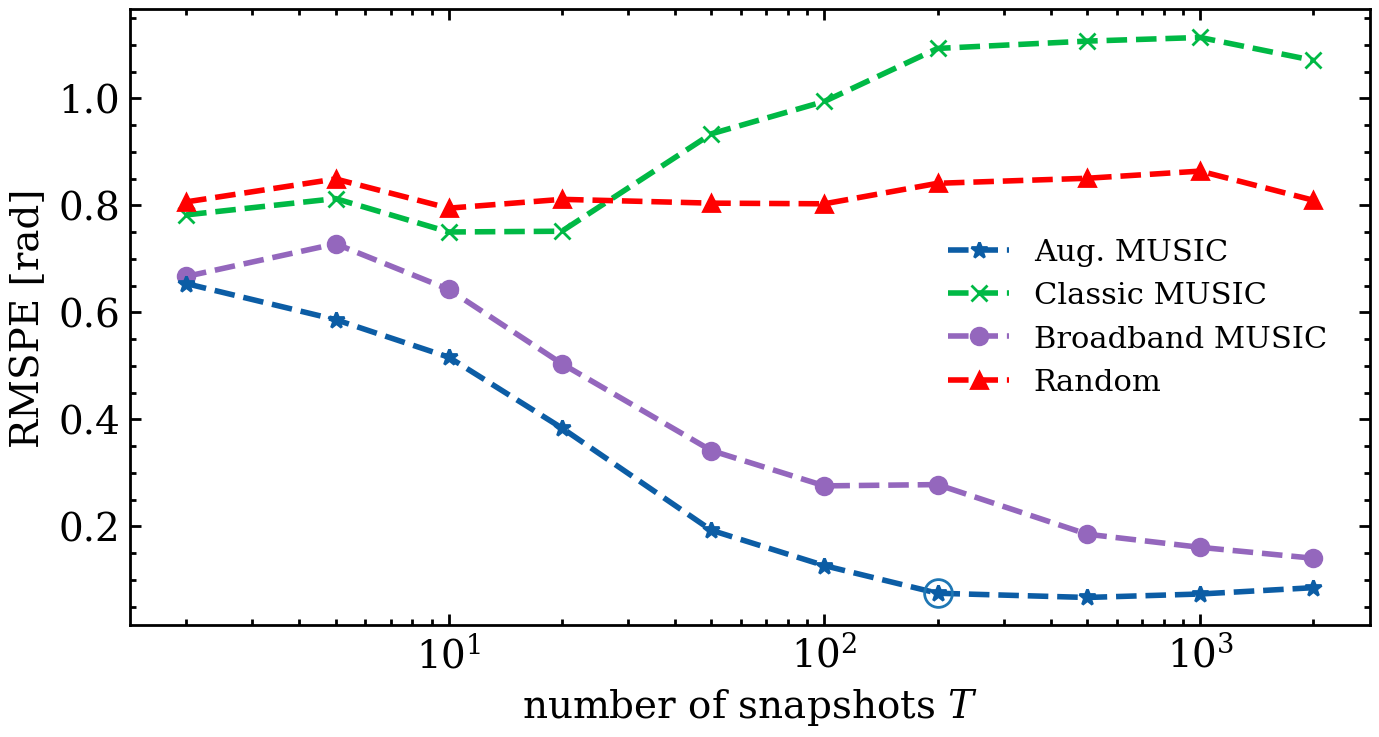}
\caption{RMSPE for known number of $D=2$ signals from Broadband Scenario 2.}
\label{fig:errVsnapsS2}
\vspace{2mm}
\centering
\includegraphics[width=1.0\columnwidth]{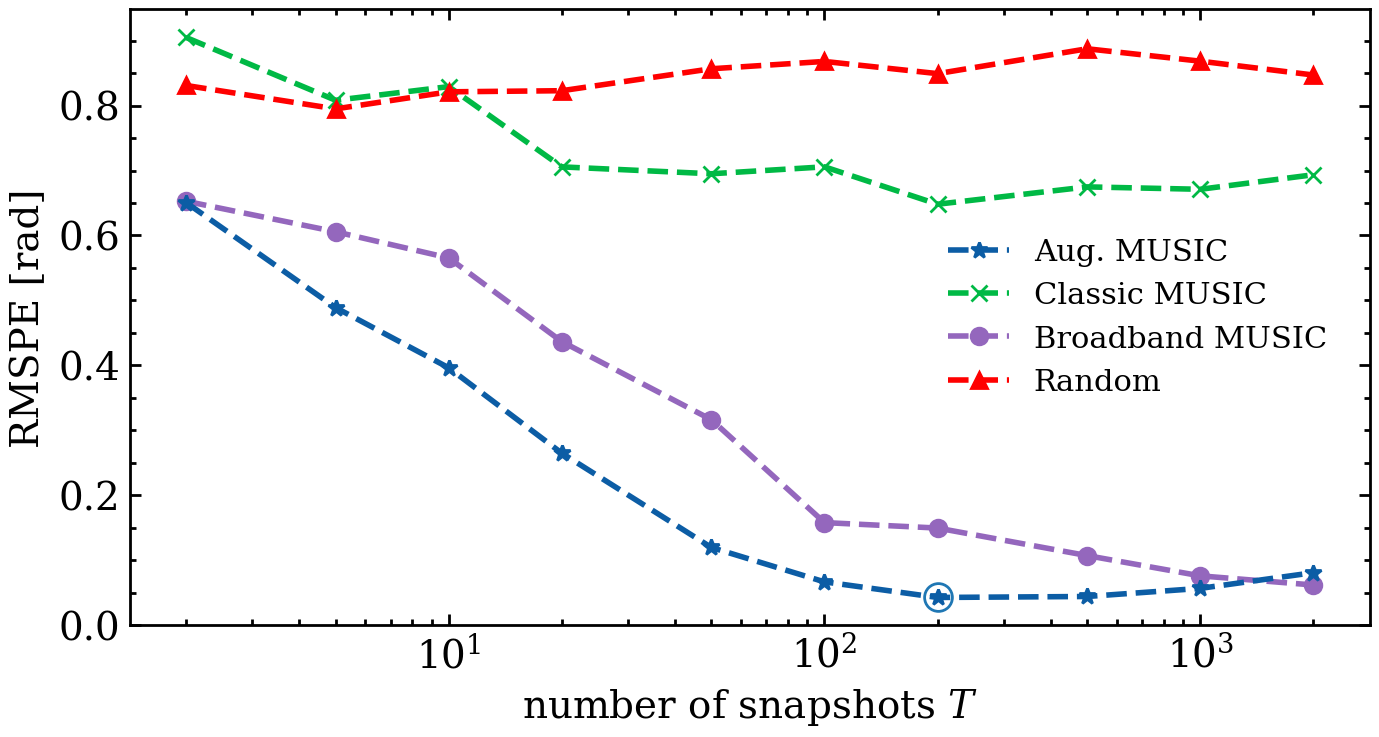}
\caption{RMSPE for known number of $D=2$ signals from Broadband Scenario 3.}
\vspace{-0.5cm}
\label{fig:errVsnapsS3}
\end{figure}
We consider the following three different signal models of the broadband signals $s_d(t)$ for $d \in \{1, ..., D\}$:
\subsubsection{Broadband Scenario 1} \label{sec:bbScen1}
A broadband signal is obtained as narrowband signals modulated on different carrier frequencies, i.e.,
\begin{equation}
    s_d(t) = \bar{s}_d \ \exp{(2 \pi \imag f_{c, d} t)},
\end{equation}
where for each $d \in \{1, ..., D\}$, both $\bar{s}_d$ and $f_{c, d}$ are randomly drawn from $\mathcal{CN}(0, 1)$ and $\mathcal{U}(f_{\min}, f_{\max})$ respectively.

\subsubsection{Broadband Scenario 2} \label{sec:bbScen2}
Broadband signals are obtained via \ac{ofdm}~\cite{weinstein2009ofdm}, which are modulated on the same carrier frequency. The signals are considered in baseband and take the following form
\begin{equation}
    s_{d, \textrm{OFDM}} = \frac{1}{K} \sum_{k = 0}^{K-1} \bar{s}_k \ \exp{(2 \pi \imag k \Delta f_d t / K)},
\end{equation}
where $\bar{s}_k \sim \mathcal{CN}(0, 1)$ is randomly drawn for each of the $K$ subcarriers and the bandwidth is $\Delta f_d = f_{\max} - f_{\min}$. 

\subsubsection{Broadband Scenario 3} \label{sec:bbScen3}
A combination of the two previous scenarios and consists of \ac{ofdm} signals modulated on different carrier frequencies
\begin{equation}
    s_d(t) =  \exp{(2 \pi \imag f_{c, d} t)} \ s_{d, \textrm{OFDM}},
\end{equation}
where $f_{c, d}$ is drawn randomly from $\mathcal{U}(f_{\min}, f_{\max} - \Delta f_d)$ to account for the signal bandwidth $\Delta f_d$.

{\bf Results:} 
Figs. \ref{fig:errVsnapsS1}, \ref{fig:errVsnapsS2}, and \ref{fig:errVsnapsS3} present the \ac{rmspe} obtained when localizing $D=2$ broadband signals from Broadband Scenario 1, 2, and 3 respectively. The number of snapshots goes as high as the sampling frequency $f_s = 2000 \Hz$ and is given logarithmically. This high number is suitable for the \ac{mb} broadband \ac{music} algorithm to achieve reliable transformation from the time domain to the frequency domain. \Ac{damusic} is again only trained for the case $T = 200$, as indicated by a circle around the marker, yet manages to perform similarly well with a higher number of snapshots. As expected, the classic narrowband \ac{music} algorithm completely fails to operate with these broadband signals, while \ac{damusic} consistently achieves the most accurate estimates, outperforming the \ac{mb} broadband \ac{music} algorithm, except for Broadband Scenario 3 with a very large number of snapshots $T > 10^3$, where \ac{damusic} trained with much shorter sequences is slightly outperformed by the \ac{mb} estimator. These results demonstrate the suitability of \ac{damusic} for coping with broadband scenarios  with a limited number of observations. 

\begin{figure}
\centering
\includegraphics[width=1.0\columnwidth]{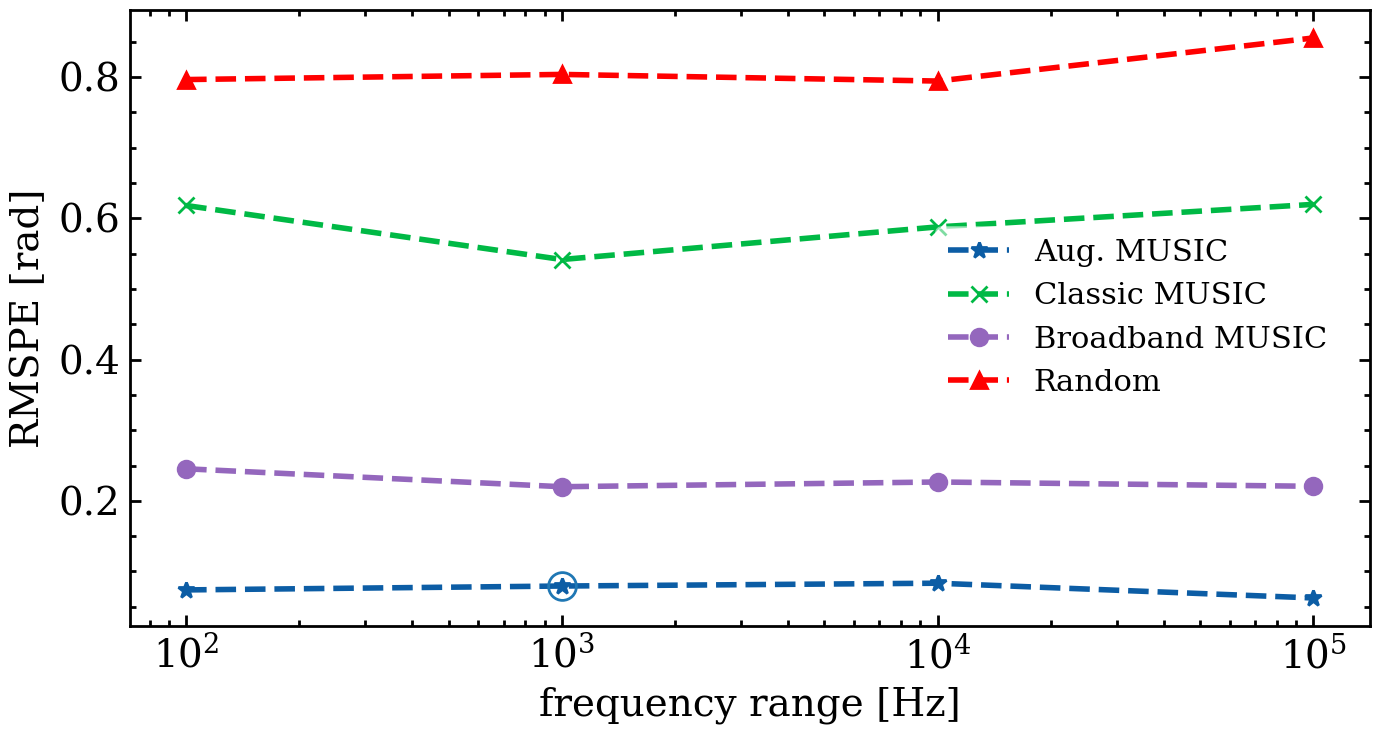}
\caption{RMSPE of varying frequency range for the carrier frequencies of Broadband Scenario 1 signals.}
\label{fig:errVfreqS1}
\vspace{2mm}
\centering
\includegraphics[width=1.0\columnwidth]{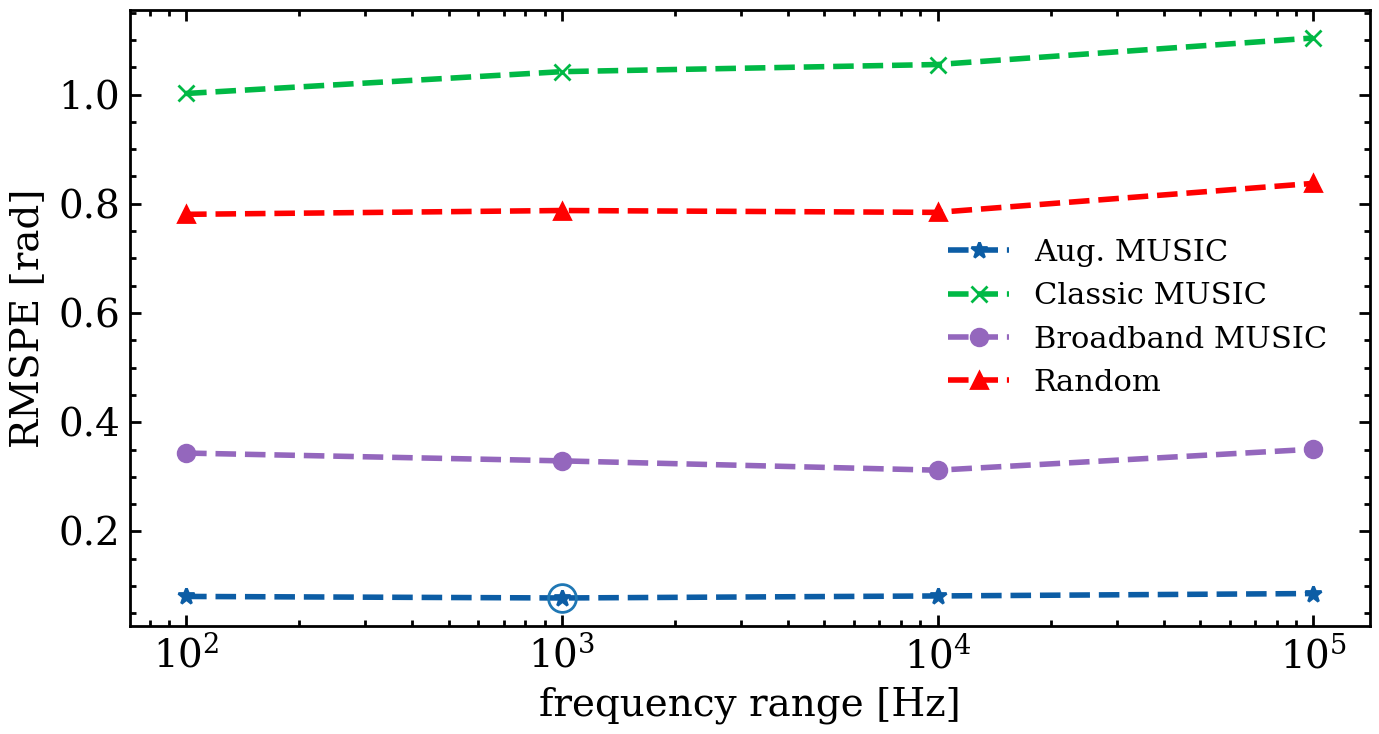}
\caption{RMSPE of varying frequency range for the bandwidth of Broadband Scenario 2 signals.}
\label{fig:errVfreqS2}
\vspace{2mm}
\centering
\includegraphics[width=1.0\columnwidth]{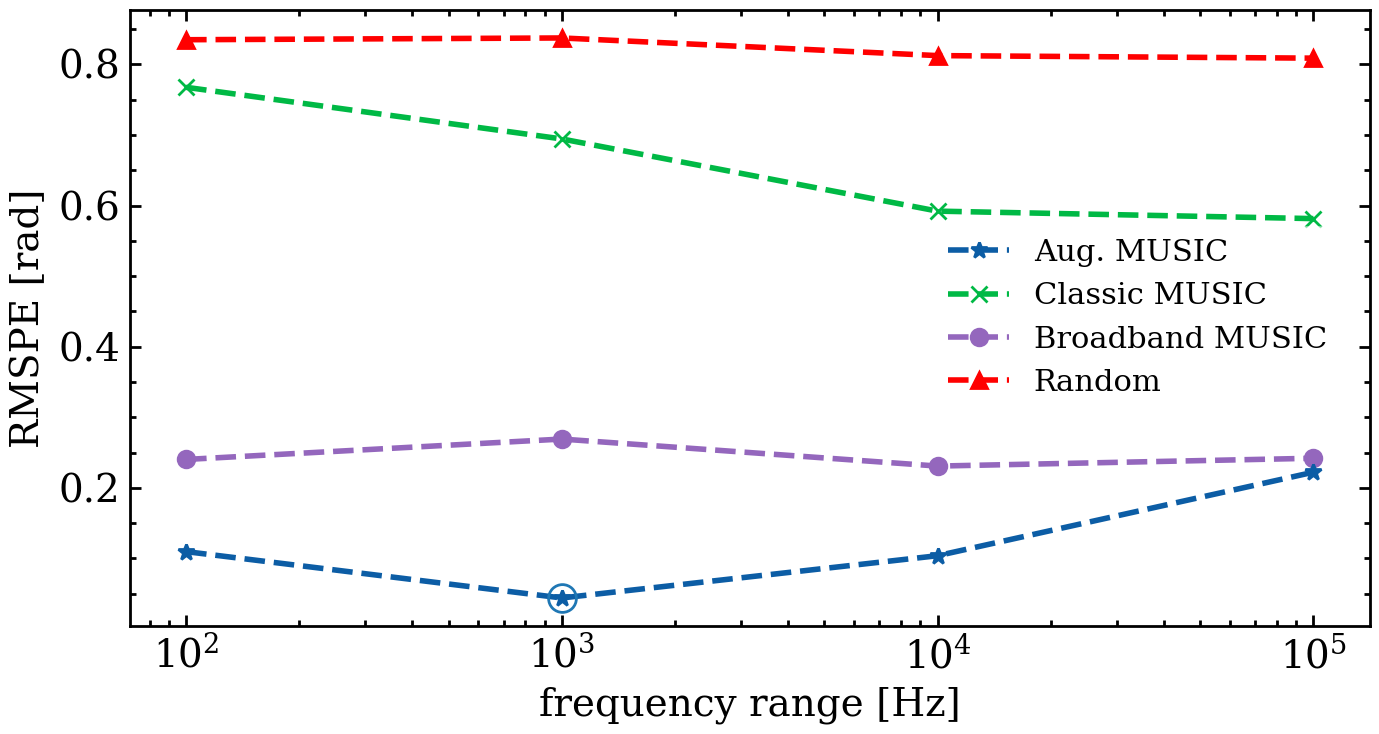}
\caption{RMSPE of varying frequency range for the carrier frequencies of Broadband Scenario 3 signals.}
\label{fig:errVfreqS3}
\end{figure}

Figs. \ref{fig:errVfreqS1}, \ref{fig:errVfreqS2}, and \ref{fig:errVfreqS3} analyze the performances with differently sized frequency ranges $[f_{\min}, f_{\max}]$. It is noted that the architecture of \ac{damusic} is almost invariant towards such scalings and manages to handle signals during inference with much larger bandwidths or modulated with higher carrier frequencies than the signals of the training data. Specifically, \Ac{damusic} is only trained for signals with carrier frequencies and bandwidths within 0 to 1000 $\Hz$ as indicated by the circle around the marker. The results in  Figs. \ref{fig:errVfreqS1}-\ref{fig:errVfreqS3} show that \ac{damusic}, whose complexity is fixed, learns to achieve the most accurate estimates, outperforming the \ac{mb} broadband \ac{music}; the latter requires an increase of the number of \acp{ifb} to overcome a larger frequency range and has a constant $10 \Hz$ per bin in the depicted results leading to a severe increase in computational complexity.

\subsection{Non-Synthetic Data: Azimuth Estimation in Seismic Arrays} \label{sec:non-synth}
In this section, we demonstrate the feasibility and the performance of  \ac{damusic} in processing non-synthetic seismic data. The seismic data was recorded by the \ac{geres} array located in the Bavarian Forest, Germany. \Ac{geres} is part of the \ac{ctbto} international monitoring system, and is a well-maintained and calibrated station. Data from \ac{geres} is continuously streamed to the \ac{idc} of the \ac{ctbto}, where it is analyzed. The array is composed of 25 vertical seismometers with a minimal distance between the sites of 124 [m] and an aperture of approximately 2.13 [km]. The seismic signal at each sensor is sampled at $40 \Hz$. Details about the \ac{geres} arrays and the exact array configuration can be found in \cite{harjes1990designAS}. 

We use data from October to December 2021, in which \ac{geres} detected arrivals for 2904 events of which 2816 were used during this analysis, 2534 for training, and 282 for testing. 
We employ sliding windows of length 100 seconds with a shift of 25 seconds around the signal arrival time as designated by the \ac{idc} analysis. 
The following parameters for each event are obtained from the \ac{idc}'s Reviewed Event Bulletin: 
the azimuth \ac{doa} angle $\theta$, the slowness value $u$, and the sensor positions $\vect{r}_1, ..., \vect{r}_M$.
Using these parameters, the steering vectors  take the following form:
\begin{equation*}
    \!\vect{a}(f,u,\psi,\alpha) \!=\! \begin{bmatrix} e^{- \imag 2 \pi f u \vect{r}_1 \vect{k}(\psi,\alpha)}, \ \dots, \ e^{- \imag  2 \pi f u \vect{r}_M \vect{k}(\psi,\alpha)}\end{bmatrix},
\end{equation*}
for some frequency $f$ and with the wave vector for certain elevation $\alpha$ and azimuth $\psi$ of interest,
\begin{equation*}
    \vect{k}(\psi,\alpha) = [\sin{\alpha} \cos{\psi}, \sin{\alpha} \sin{\psi}, \cos{\alpha}].
\end{equation*}
%
\begin{table}
    \centering
    \caption{\Ac{rmspe} in [rad] of different \ac{doa} estimation algorithms for seismic data.}
    \label{tab:resultsReal}
    \begin{tabular}{ | c | c | c | c | c | c |} 
        \hline
       & \thead{DA-MUSIC} & \thead{Broadband\\MUSIC} & \thead{Classic\\MUSIC} & \thead{Beam-\\former} & \thead{Random} \\
        \hline
         \thead{RMSPE} & \textbf{0.6269} & 1.0075 & 1.2475 & 0.9383 & 1.7097 \\
        \hline
    \end{tabular}
\end{table}
We compare \ac{damusic} with following \ac{doa} estimators for the setting settings  $\alpha=-\pi/4$ and $f=1 \Hz$:
\begin{itemize}
    \item Broadband \ac{music}, corresponding to an incoherent broadband extension of \ac{music} \cite{yoon2006doaest}, and instead of using the constant $f=1 \Hz$ 
    it utilizes 10 \ac{ifb} with frequencies in [0, 20] $\Hz$.
    \item The classic MUSIC algorithm, implemented in its narrowband format as described in Section~\ref{sec:music} with additionally filtering the measurements with an experimentally calibrated low-pass filter, allowing only frequencies within [0, 10] $\Hz$ to pass.
    \item A conventional beamformer~\cite{bartlett1948beamformer}.
    \item Choosing a \ac{doa} angle at random.
\end{itemize}
The results, reported in Table~\ref{tab:resultsReal}, show that \Ac{damusic} manages to outperform the \ac{mb} estimators by not only focusing the frequency component, but also by concurrently focusing the interdependent elevation angle to receive a stable azimuth estimation. On the other hand, the \ac{mb} algorithms require further knowledge of the elevation and frequency at hand to operate reliably. While the errors in Table~\ref{tab:resultsReal} may appear to be relatively large, it is noted that the average error achieved via expert analysis reported by the \ac{idc} Reviewed Event Bulletin is 0.4243 [rad]. This indicates the ability of \ac{damusic} to achieve comparable results and to notably outperform the \ac{mb} estimators while operating with simplified and approximated model parameters.
%




\section{Conclusions} \label{sec:conclusion}
We presented a hybrid \ac{mb}/\ac{dd} implementation of the \ac{music} algorithm for \ac{doa} estimation. The proposed \ac{damusic} was shown to mitigate some of the limitations and drawbacks of the classic method. \ac{damusic} is operable with an unknown number of sources and with broadband signals while being adaptable to various scenarios  
and robust towards severe mismatches in the array geometry. 
The proposed hybrid \ac{mb}/\ac{dd} approach provides a viable alternative in both low and high snapshot domains and shows a remarkable resolution capability compared to both \ac{mb} and \ac{dd} benchmarks in various settings.


\section{Acknowledgment}
We thank Dr. Yochai Ben Horin for constructive and valuable joint discussions on the seismic data, and for providing the data.


\bibliography{IEEEabrv, refs}
\bibliographystyle{IEEEtran}

\end{document}